\documentclass[%
 prl, twocolumn,
 amsmath,amssymb,
 aps,
]{revtex4-1}

\usepackage{graphicx}
\usepackage{dcolumn}
\usepackage{bm}
\usepackage{xcolor}
\usepackage{xr}
\usepackage{bm}
\usepackage{hyperref}
\usepackage{enumitem}
\usepackage{accents}
\hypersetup{colorlinks,linkcolor={blue!90!black},citecolor={blue!90!black},urlcolor={blue!90!black}}

\DeclareMathAlphabet\mathbfcal{OMS}{cmsy}{b}{n}

\usepackage{amsmath}
\usepackage{mathtools}

\newcommand\filledcirc{{\color{cyan}\bullet}\mathllap{\circ}}
\usepackage{siunitx}
\newcommand{\etal}{\textit{et al.}}

\begin{document}

\title{Biofilm Growth Under Elastic Confinement}

\author{George T. Fortune$^1$, Nuno M. Oliveira$^{1,2}$, and Raymond E. Goldstein$^1$
  \email{R.E.Goldstein@damtp.cam.ac.uk}}
\affiliation{$^1$Department of Applied Mathematics and Theoretical 
Physics, Centre for Mathematical Sciences, University of Cambridge, Wilberforce Road, Cambridge CB3 0WA, 
United Kingdom \\
$^2$Department of Veterinary Medicine, University of Cambridge, Madingley Road, Cambridge CB3 0ES, 
United Kingdom}%
\def\nuno#1{\textcolor{red}{\normalsize #1}}
\def\george#1{\textcolor{blue}{\normalsize #1}}
\date{\today}

\begin{abstract}
Bacteria often form surface-bound communities, embedded in a self-produced extracellular matrix, called biofilms. Quantitative 
studies of their growth have typically focused on unconfined expansion above solid or semi-solid surfaces, leading to exponential 
radial growth. This geometry does not accurately reflect the natural or biomedical contexts in which biofilms grow in confined spaces. Here we 
consider one of the simplest confined geometries: a biofilm growing laterally in the space between a solid surface and an overlying 
elastic sheet. A poroelastic framework is utilised to derive the radial growth rate of the biofilm; it reveals an additional 
self-similar expansion regime, governed by the stiffness of the matrix, leading to a finite maximum radius, consistent with our experimental 
observations of growing \textit{Bacillus subtilis} biofilms confined by PDMS. 
\end{abstract}

\maketitle

Bacterial biofilms are microbial accretions, enclosed in a self-produced polymeric extracellular matrix \cite{Bjarnsholt11}, which adhere 
to inert or living surfaces. A biofilm gives the individual cells a range of competitive advantages, such as increased resistance to chemical attack.
Since the popularisation in the mid 1600s of the light microscope as a tool to study problems in biology \cite{Hooke1665,Saraf84}, 
observations of groups of bacteria on surfaces have been amply documented \cite{Wimpenny2000}, most notably by van Leeuwenhoek 
in his dental plaque \cite{vanL}. Yet, it is only in the last few decades with the development of new genetic and molecular techniques 
that the complexity of these communities has been appreciated and biofilm formation has been recognised as a
regulated developmental process in its own right \cite{Costerton1994, OToole2000}. 

Biofilm formation is common across a wide range of organisms in the archaeal and bacterial domains of life, on almost all types of 
surfaces \cite{Lopez10}. Cells attach to a surface and form micro-colonies through clonal growth. These then 
grow and colonise their surroundings through twitching motility \cite{Bjarnsholt11}. A central research focus has been understanding these growth 
dynamics.  Building on important work on osmotically-driven spreading \cite{Seminara12}, a biofilm has often been modelled as a viscous, 
Newtonian fluid mixture (nutrient rich water and biomass), neglecting the matrix elasticity. The effects of surface tension \cite{Tam19}, 
osmotic pressure \cite{Srinivasan19}, and the interplay between nutrients, cell growth, and electrical signaling in response to metabolic 
stress have all been studied recently \cite{Martinez-Corral19}. 

While previous analyses have focused on the experimentally tractable cases of unconfined and unsubmerged 
biofilms \cite{Seminara12,Tam19,Srinivasan19,Martinez-Corral19}, they do not accurately reflect the conditions 
in which many biofilms grow; they thrive in confined micro-spaces \cite{Kempf19} between flexible elastic boundaries 
such as vessel walls or soil pores \cite{Conrad18}, and indeed in the human body, where they account for over $80\%$ of 
microbial infections \cite{Khatoon18}. Biofilms are difficult to treat with antibiotics, being thousands of times more resistant 
than the constituent microorganisms in isolation \cite{Oppenheimer13} due to a range of mechanical and biological processes 
\cite{Ciofu17,Stewart02}. The recent rapid growth in the use of implantable biomedical devices (stents, catheters, and cardiac 
implants) has brought with it a large increase in associated biofilm infections \cite{Arciola18} since artificial surfaces require 
much smaller bacterial loads for colonisation than the corresponding volume of native tissue ($\approx 10^{-4}$ as much \cite{Nowakowska14}). 

\begin{figure}[b]
	\centering
	\includegraphics[trim={0 0cm 0 0cm}, clip, width=0.45\textwidth]{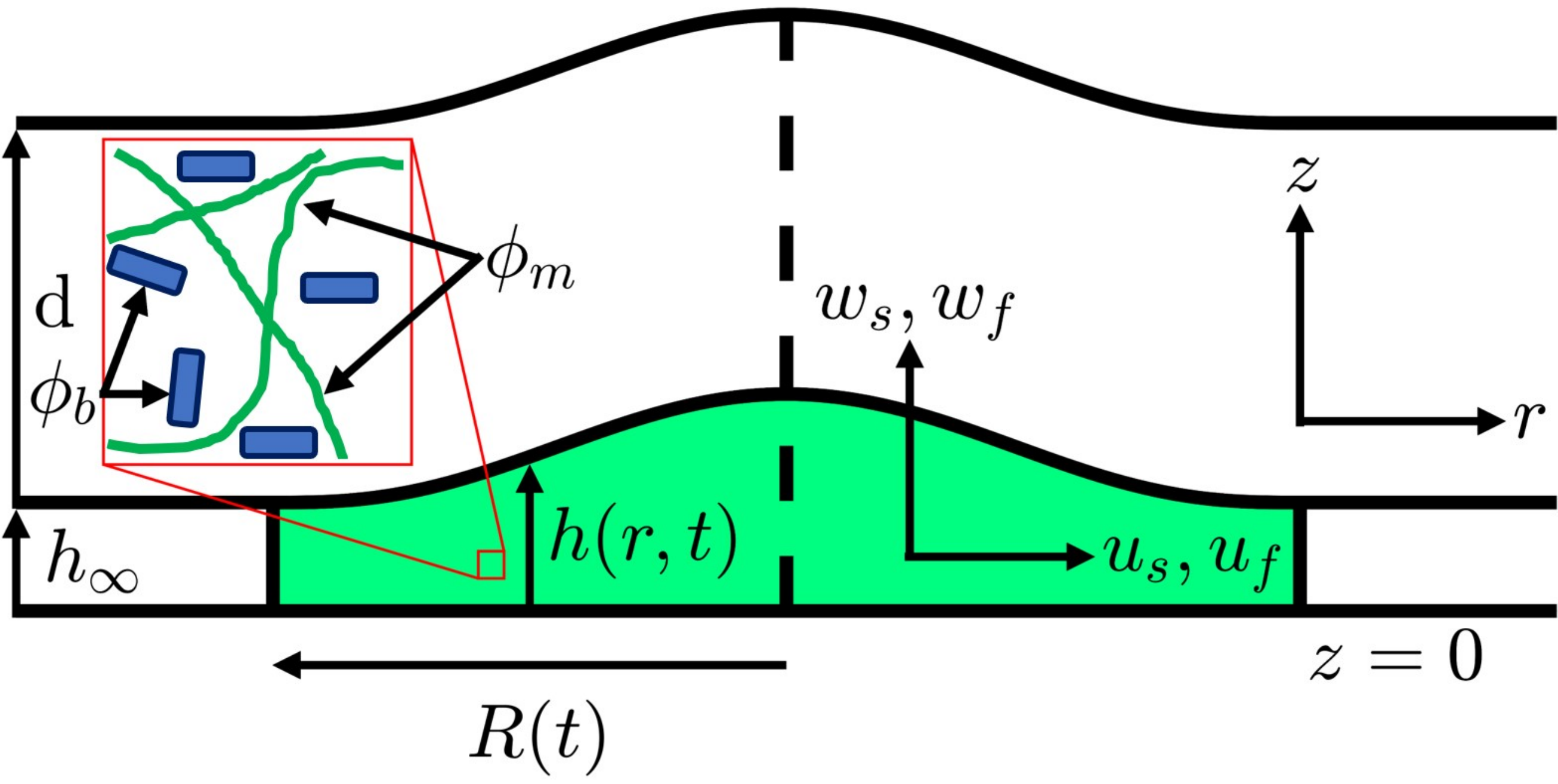}
	\caption{Schematic of a confined biofilm. An axisymmetric biofilm (green) grows between a rigid surface at $z = 0$ and 
	an elastic sheet at $z = h$, with undeformed gap height $h_{\infty}$. Inset: the biomass is a mixture of bacterial cells 
	(blue, volume fraction $\phi_b$) and extracellular matrix (green,volume fraction $\phi_m$). The pore-averaged velocities 
	of the solid and fluid phases are denoted by $\boldsymbol{u_s} = (u_s, \, w_s)$ and $\boldsymbol{u_f} = (u_f, \, w_f)$.}
	\label{fig:1}
\end{figure}

Here we develop the simplest model for a confined biofilm, using a poroelastic framework to obtain a system of equations 
describing its expansion dynamics. We find an analytic similarity solution for the biofilm height and radius, together with the 
vertically averaged biomass volume fraction. Consistent with experimental observations on growing \textit{Bacillus subtilis} 
biofilms described here, unlike unconfined biofilms whose radius grows exponentially, the balance between elastic stresses and 
osmotic pressure difference across the interface implies an additional possible growth regime where within a shallow layer lubrication 
assumption, confined biofilms have a maximum radius at long times. The transition between regimes is governed by the stiffness 
of the matrix.

We consider a bio-mechanical system in which bacteria grow and divide, converting 
nutrient-rich fluid into biomass and thus inducing a flow of biomass outwards from the biofilm centre. This flow is resisted 
by elastic stresses within the extracellular matrix (ECM), while the biofilm height dynamically adjusts to ensure conservation of 
normal stress across the overlying elastic sheet. An influx of water assures volume conservation.
Illustrated in Fig. \ref{fig:1}, an axisymmetric biofilm of thickness $h(r,t)$, radius $R(t)$ and biomass volume $V$ rests on 
an impermeable flat plate at $z = 0$ and grows below an elastic sheet of thickness $d = \mathcal{O}(R)$ and bending 
modulus $B = Ed^3 / 12 (1 - \nu^2)$, where $E$ and $\nu$ are the Young's modulus and Poisson's ratio of the sheet. We examine 
the simplest biofilm composition, a mixture of bacteria (volume fraction $\phi_b$), sugar-rich secreted polymeric ECM 
(volume fraction $\phi_m$), and nutrient-rich water (modelled as a low viscosity Newtonian fluid \cite{Seminara12} with 
dynamic viscosity $\mu_{f}$ and volume fraction $1 - (\phi_m + \phi_b) \equiv 1 - \phi$), under the assumption that 
$\phi_m \ll \phi_b$ \cite{Seminara12}. For theoretical simplicity, we assume that the biomass volume fraction $\phi$ is 
independent of $z$, so $\partial\phi/\partial z=0$.

We denote the pore-averaged velocity and stress tensor of the solid and liquid phases by 
$\{ \boldsymbol{u_s} = (u_s, w_s) \, , \, \boldsymbol{\sigma_{s}} \}$ and 
$\{ \boldsymbol{u_f} = (u_f, w_f) \, , \, \boldsymbol{\sigma_{f}} 
\approx -p\boldsymbol{I} \, \}$ \cite{Seminara12} respectively, where $p$, $\Pi$ and $\tilde{p}$ are the pore, 
osmotic, and bulk pressures (with $\tilde{p} = p + \Pi$ \cite{Peppin05}). Since the vertical deflection of the 
sheet $\Delta d = \mathcal{O}(h)$ is small compared to its thickness $d$, we ignore stretching and model 
it as a thin elastic beam with radius of curvature $\tilde{R} \gg \{d, \, h \}$ and 
surface tension $\gamma$ against the biofilm. We neglect gravity, assume that nutrient concentrations across the 
biofilm are constant, and take the biomass growth rate $g$ to have the saturating form
\begin{equation}
    g = \frac{1}{T_{D}}\left(\frac{c}{c + c_{\text{half}}}\right), \label{eq:growth}
\end{equation}
independent of position, where $T_D$ is the doubling time (typically hours), $c$ is the 
concentration of a limiting nutrient and $c_{\text{half}}$ is that for half-maximum growth rate. Both $c$ and hence $g$ 
are taken to be constant in light of our experiments, introduced below, in which there is an external flow 
that ensures homogeneity.  Conserving mass in both the solid and fluid phases gives
\begin{subequations}
\begin{align}
\frac{\partial \phi}{\partial t} + \boldsymbol{\nabla \cdot} (\phi \boldsymbol{u_s}) &= g \phi, \label{eq:massconservation1}\\
-\frac{\partial \phi}{\partial t} + \boldsymbol{\nabla \cdot}((1-\phi)\boldsymbol{u_f}) &= - g \phi. \label{eq:massconservation2}
\end{align} \label{eq:massconservation}
\end{subequations}

Defining the Terzaghi effective stress tensor as 
$\boldsymbol{\sigma} = \phi(\boldsymbol{\sigma_s} - \boldsymbol{\sigma_f})$ \cite{Wang01}, 
momentum balance yields
\begin{equation}
\boldsymbol{\nabla \cdot \sigma} = \boldsymbol{\nabla}p. \label{eq:terzaghi}
\end{equation}
To model $\boldsymbol{\sigma}$, we deviate from prior work that assumed a Newtonian fluid by adopting a poroelastic framework that
incorporates the elasticity of the ECM. In this picture, $\boldsymbol{\sigma}$ obeys the elastic constitutive law
\begin{equation}
    \sigma = \sigma \left( \boldsymbol{\nabla \xi} \right),
\end{equation}
where $\boldsymbol{\xi} = (\xi, \zeta)$, the deformation vector of the medium away from a reference state, is related to 
the biofilm phase velocity through $\boldsymbol{u_s} = \left(\partial_t + \boldsymbol{u_s \cdot \nabla}\right)\boldsymbol{\xi}$. 
Little utilised in the study of biofilms, it is a common approach in many problems containing elasticity in geophysics 
(hydrology subsidence and pumping problems \cite{Hewitt15,Gibson70} or industrial filtration \cite{Barry97}) and  biological 
physics (cell cytoplasm \cite{Charras09}). Here, we consider the simplest case, where $\boldsymbol{\sigma}$ obeys the linear 
constitutive law
\begin{equation}
    \boldsymbol{\sigma}(\boldsymbol{\nabla \xi}) = \left(K - \frac{2G}{3}\right)(\boldsymbol{\nabla \cdot \xi})\boldsymbol{I} 
    + G(\boldsymbol{\nabla \xi} + \boldsymbol{\nabla \xi}^T),
\end{equation}
where $K$ and $G$ are the effective bulk and shear moduli of the biofilm respectively, assumed constant. 
As in \cite{Hewitt15}, $K$ and $G$ are properties of the whole biofilm rather than just the ECM. We prescribe explicitly 
the general form for the horizontal velocity of the solid phase,
\begin{equation}
    u_s = \frac{r}{R}\frac{\partial R}{\partial t} u_{0}\left( \frac{z}{h} \right), \label{eq:govern5}
\end{equation}
where $u_{0}$ is the $z-$dependent part of $u_s$. We take 
\begin{equation}
u_{0} = \frac{6z(h-z)}{h^2}, \label{eq:govern6}
\end{equation}
since this is the simplest functional form obeying no-slip boundary conditions at $z = 0$ and $z = h$ as well 
as $\langle u_{0}\rangle = 1$. However, as shown below, we find a solution independent of the exact form for 
$u_{0}$. Global volume conservation gives $\partial R/\partial t$ while $r/R$ sets a simple linear radial dependence, 
ensuring that $u_s = 0$ at $r = 0$. As for $u_{0}$, tweaking this radial dependence does not qualitatively 
change the resulting dynamics of the system.

In contrast, vertical flow is governed by pressure gradients induced both by the upper confinement and by elastic stresses 
in the extracellular matrix. We invoke Darcy's law for flow within the matrix, giving 
\begin{equation}
(1 - \phi)(w_s - w_f) = \frac{\kappa}{\mu_f} \frac{\partial p}{\partial z}. \label{eq:govern7}
\end{equation}
where $\kappa = \kappa(\phi)$ is the effective biofilm permeability with characteristic permeability scale $\kappa_0$. 
The osmotic pressure away from equilibrium $\Pi(\phi)$ is taken to be that ofFlory Huggins theory \cite{Flory53}, with
interaction parameter $\chi\simeq 1/2$ so there is no demixing \cite{Winstanley11}.  Assuming that the matrix solid 
fraction $\beta = \phi_m/\phi \ll 1$ is constant across the biofilm, the osmotic pressure is \cite{pi_caveat}
\begin{equation}
\Pi = \frac{k_B T}{3\nu_0}\left(\frac{\phi_m}{1-\phi} \right)^3, \label{eq:osomotic}
\end{equation}
a function of thermal energy $k_{B}T$ and $\nu_0$, the effective volume occupied by one monomer of matrix. 
Since the matrix consists of many different substances, notably sugars, proteins and DNA, we estimate $\nu_0$ by the 
volume occupied by one sugar monomer. This term is subdominant in the analysis below, and thus does not appear 
in the interior ($r \leq R$) solutions \eqref{eq:interiorhphi} - \eqref{eq:interiorR}. We close this system of equations 
with a set of vertical boundary conditions, given in the Supplementary Material \cite{suppmat}. 

The analysis exploits two separations of scales: (i) the initial radius of the confined 
biofilm $R_0 = R(t = 0)$ is much greater than the initial 
height $H_0 = h(r = 0, t = 0)$, a lubrication approximation, and (ii) the growth time scale $1/g$ is much 
larger than the poroelastic equilibration time $\mu_f H^2_0 / \kappa_0 P_0$. We nondimensionalise the equations anisotropically using these 
length scales, denote the vertically averaged form of a function $f$ by $\langle f \rangle = h^{-1}\int^{h}_{0} f \, dz$, 
and define $\varphi = \langle \phi \rangle$, $v_s=\langle u_s \rangle$, $k = \langle \kappa \rangle$, 
$\mathcal{P} = p/P_0$ and
\begin{equation}
    \rho = \frac{r}{R(0)}, \ \ \tau = gt, \ \ {\mathcal{R}} = \frac{R(t)}{R(0)}, 
   \ \ {\cal H} = \frac{h(r,t)}{h(0,0)}.
 \end{equation}
Keeping only leading-order terms in $\epsilon=H_0/R_0$ \cite{suppmat}, the model reduces to coupled PDEs for 
the height ${\cal H}(\rho,\tau)$ and depth-averaged biomass fraction $\varphi(\rho,\tau)$ as functions of radial 
distance $\rho$ and time $\tau$. The horizontal pressure gradient adjusts to one of three possible modes
\begin{equation}
    \frac{\partial \mathcal{P}}{\partial \rho} = \left\{ 0, \, \frac{C_1}{\rho}, \frac{C_2}{\rho^2}, \right\} \label{eq:pressurecondition}
\end{equation}
where $C_1$ and $C_2$ are constants and the dominant contribution to the pressure $\mathcal{P}$ arises from the 
bending stresses imposed from the upper elastic sheet,
\begin{equation}
{\mathcal P} = \nabla^4 {\cal H}.
\end{equation}
The depth-integrated biomass fraction $\varphi\mathcal{H}$ satisfies a conservation law of the form 
$\partial (\varphi{\cal H})/\partial \tau=-{\bm \nabla}\cdot {\mathbfcal J}_{\cal \varphi} + {\mathbfcal S}$,
\begin{equation}
      \frac{\partial}{\partial \tau}(\varphi{\cal H}) = -\frac{1}{\rho}
      \frac{\partial}{\partial \rho}\left(\rho v_s \varphi{\cal H}\right) + \varphi{\cal H}.
      \label{eq:finalphiequation}
\end{equation}
Thus, $\varphi \mathcal{H}$ grows exponentially from the source term ${\mathbfcal S} = \varphi{\cal H}$, while subject to 
radial advection at speed $v_s({\cal H},\mathcal{R})$ from the flux term ${\mathbfcal J}_{\cal \varphi}$. The system is 
closed with a set of boundary conditions, deriving the boundary conditions for ${\cal H}$ at the biofilm interface by 
extending the framework outside the biofilm to the whole domain and imposing far field boundary conditions \cite{suppmat}.   
\begin{figure}
    \centering
    \includegraphics[width=\columnwidth]{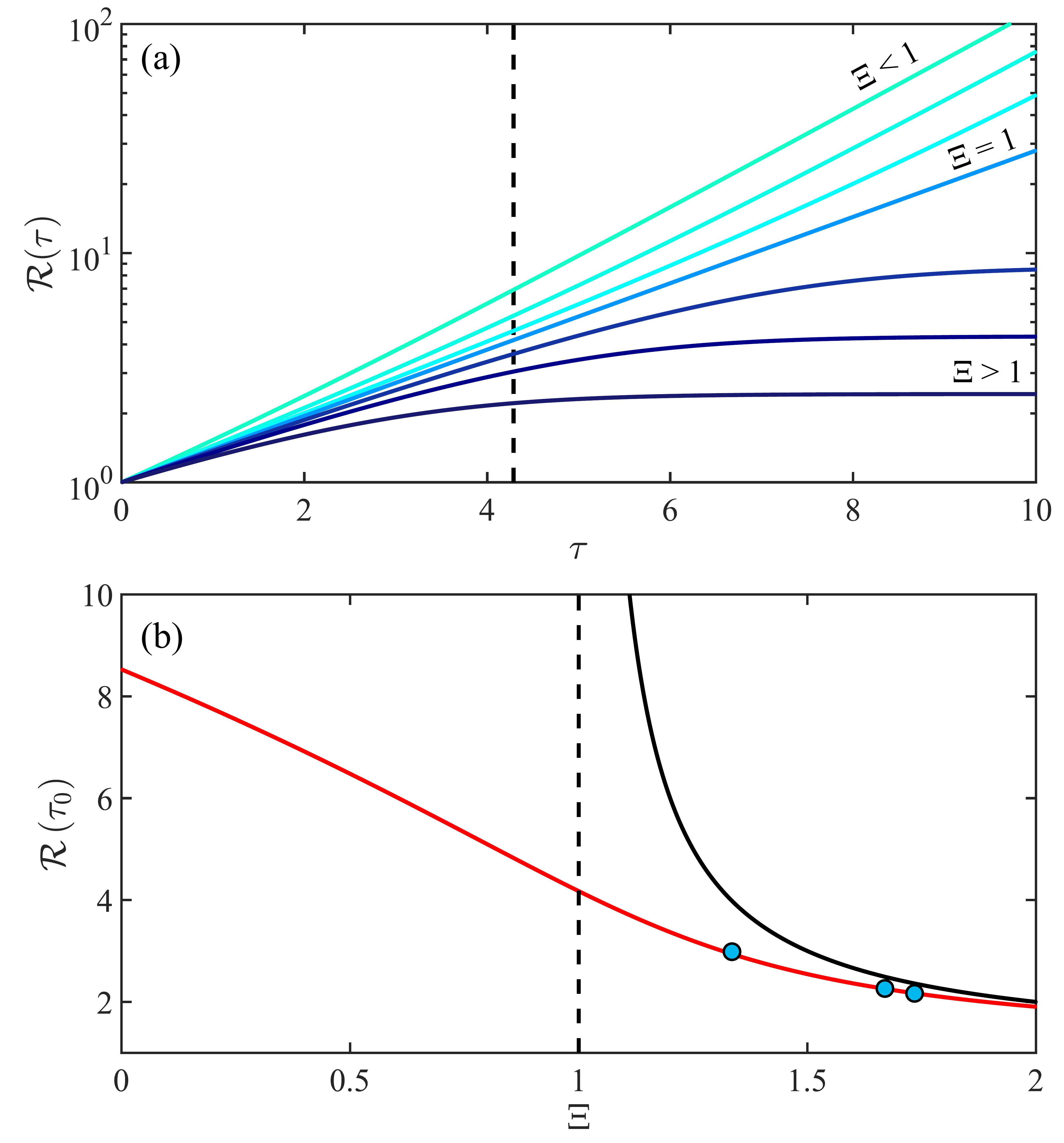}
    \caption{Growth dynamics of confined biofilms according to the poroelastic model. (a) The 
    scaled biofilm radius $\mathcal{R}$ as a function of scaled time in a semilogarithmic plot, 
    for $\Xi \in [0.4,0.75,0.91,1,1.13,1.3,1.7]$. Darker colours denote larger $\Xi$. 
    (b) Biofilm radius at a fixed $\tau_0$ (dashed vertical line in (a)) as a function of $\Xi$, both 
    numerically (\textcolor{red}{\textbf{-}}) and experimentally 
    ($\filledcirc$), and numerically for $\tau_0 \rightarrow \infty$ (\textcolor{black}{\textbf{-}}).
	}
    \label{fig:2}
\end{figure}
In the mode zero case when the horizontal pressure gradient is zero, Eqs. (\ref{eq:pressurecondition})-(\ref{eq:finalphiequation}) 
admit the interior ($\rho \leq \mathcal{R}$) solutions
\begin{subequations}
\begin{align}
    {\cal H} &= e^{\tau}{\mathcal{R}}^{-2}f(\rho/{\mathcal{R}}), \\
    \varphi_0 &= \varphi_0(\rho/{\mathcal{R}}),
\end{align} \label{eq:interiorhphi}
\end{subequations}
where 
\begin{equation}
    f(x) = 1 - \left(1 - m_0\right)x^2,
\end{equation}
the incline ratio
\begin{equation}
    m_0 = \frac{h(r = R(0),t = 0)}{h(r = 0,t = 0)}
    \label{hinfdefine}
\end{equation}
is a measure of the initial flatness of the biofilm, 
$\varphi_0(\rho) = \varphi(\rho,\tau = 0)$ is set from the initial conditions and we 
have utilized the vertically-averaged 
boundary conditions \cite{suppmat} and the initial 
conditions $\mathcal{H}\left(\rho = 0, \, \tau = 0\right) = \mathcal{R}\left(\tau = 0\right) = 1$ 
and $\mathcal{H}\left(\rho = 1, \, \tau = 0\right) = m_0$. The form of \eqref{eq:interiorhphi} guarantees that
the total biomass $\int d\rho \rho\, {\cal H}\varphi$ grows as ${\rm e}^{\tau}$. We obtain $\mathcal{R}(\tau)$ 
as the solution of the cubic equation 
\begin{equation}
    e^{- \tau} \mathcal{R}^3 + \mathcal{R}(\Xi - 1) - \Xi = 0, \label{eq:Rgoverningequation}
\end{equation}
where the single free parameter is 
\begin{equation}
    \Xi = \frac{\xi_0 m_0}{\zeta_0}\frac{K + G/3}{K + 4G/3} = \frac{\Psi}{2(1 - \nu_b)}. \label{eq:interiorR}
\end{equation}
Derived in \cite{suppmat}, $\Psi = \xi_0 m_0 / \zeta_0$ is a measure of the initial ratio between horizontal and 
vertical stress gradients in the biofilm while $\nu_b$, the effective Poisson's ratio of the ECM, 
is a measure of how stiff the biofilm is (stiffer biofilms have lower $\nu_b$). The radial expansion of the biofilm is 
mediated by a balance at the biofilm edge between horizontal and vertical elastic deformation in the biofilm 
(the $\Xi$ and $e^{-\tau}\mathcal{R}^3$ terms, respectively, in \eqref{eq:Rgoverningequation}) and the 
osmotic pressure difference across the biofilm interface (the $\mathcal{R}(\Xi - 1)$ term). 

For general $\Xi$ and $\tau$, this equation does not always admit an analytic solution and is 
solved numerically \cite{suppmat}. Figure \ref{fig:2}(a) plots the temporal evolution of 
$\mathcal{R}$ for a range of different values of $\Xi$. Figure \ref{fig:2}(b) explores this further, 
choosing a fixed observation time $\tau_0$ and plotting $\mathcal{R}(\tau_0)$ as a function of $\Xi$.
Two clear regimes emerge. If $\Xi < 1$, 
the first and second terms 
in \eqref{eq:Rgoverningequation} dominate in a balance between stresses caused by the vertical elastic deformations 
and the osmotic pressure difference, leading to a limit on vertical expansion.  The 
biofilm then spreads with exponential radial growth \cite{Seminara12}, with 
$\mathcal{R} \rightarrow \left( 1 - \Xi \right)^{1/2} e^{\tau/2}$ as $\tau \rightarrow \infty$.  
If $\Xi > 1$ (the dark blue curves in figure \ref{fig:2}(a)), 
the second and third term in \eqref{eq:Rgoverningequation} are dominant, 
giving a balance between stresses caused by horizontal elastic deformations and 
the osmotic pressure difference that limits horizontal expansion.
The radius at intermediate times exhibits power-law growth before slowing down to reach a maximum 
$\mathcal{R}(\infty) = \Xi/(\Xi - 1)$, when the shallow layer approximation is still valid.
In the special case $\Xi = 1$, the osmotic pressure difference 
across the interface is zero, leading to a balance between horizontal and vertical elastic stresses.
As shown in Fig. \ref{fig:2}(a), the system exhibits transitional exponential growth, with $\mathcal{R} = e^{\tau/3}$,
but this state is not stable; curves with $\Xi$ just above and below unity will veer off eventually to tend to 
a constant radius or to the faster $e^{\tau/2}$ growth law.

\begin{figure}[t]
	\centering
	\includegraphics[clip, width=0.98\columnwidth]{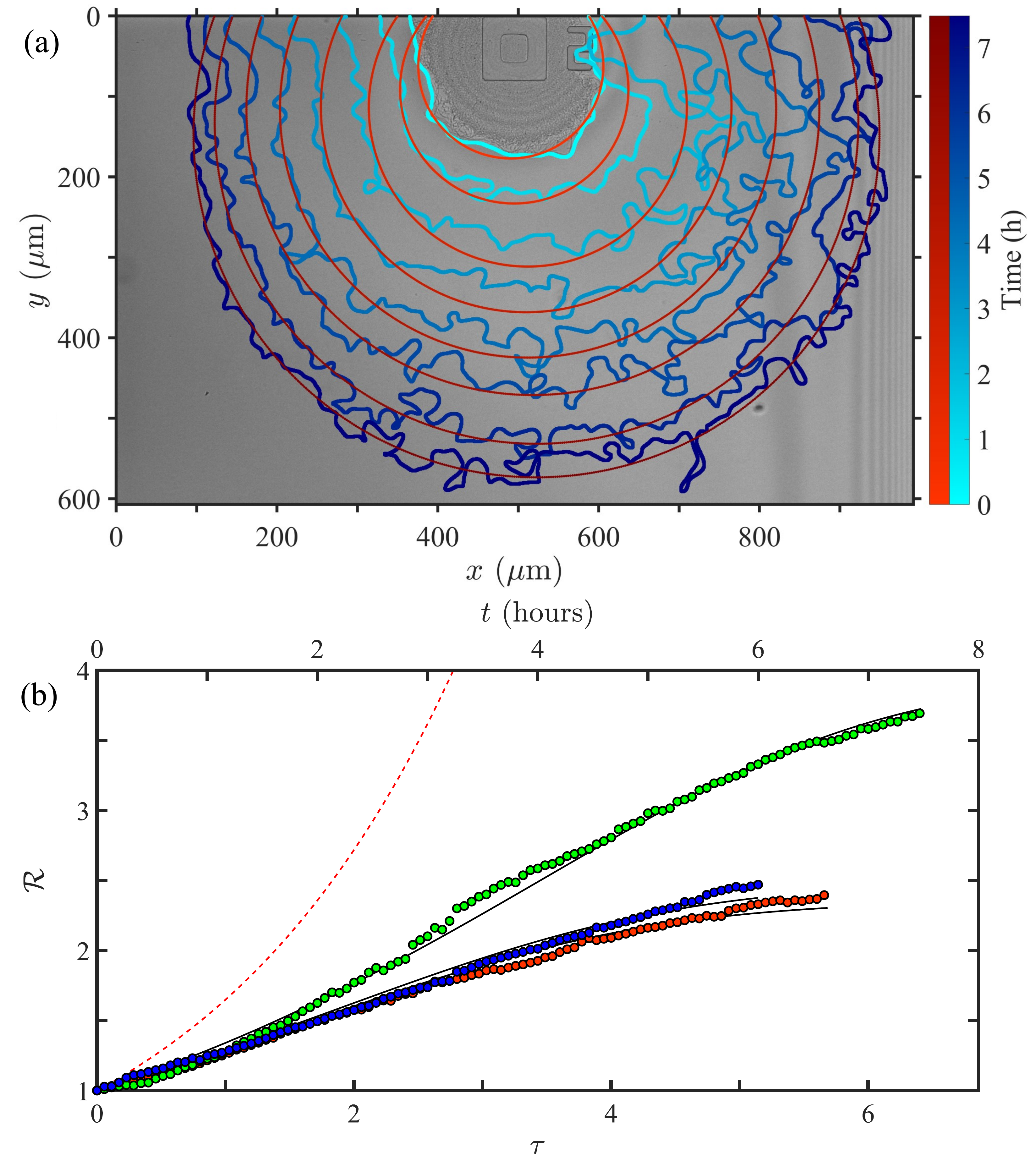}
	\caption{Experimental growth of {\it B. subtilis} biofilms under confinement by a PDMS sheet. 
	(a) Montage plot, superimposed on image of the initial biofilm, 
	showing the temporal evolution of the biofilm boundary (blue curves; darker colors denote later times) and 
	fitted circles (red). (b) Scaled biofilm radius $\mathcal{R}$ against scaled time for $3$ experiments 
	(${{\color{blue}\bullet}\mathllap{\circ}}$ , ${{\color{red}\bullet}\mathllap{\circ}}$ , ${{\color{green}\bullet}\mathllap{\circ}}$) compared to fitted dynamics from model (\textcolor{black}{\textbf{--}})  Dashed red curve is 
	${\cal R}(\tau)={\rm e}^{\tau/2}$ expected for growth at constant thickness.
	}
	\label{fig:3}
\end{figure}

We  performed experiments on the growth of biofilms confined by polydimethylsiloxane (PDMS), the
results of which can be compared directly to the model developed above. 
The methodology follows existing protocols \cite{Martinez-Corral19,Liu15,Humphries17} developed to 
understand the growth of focal (and submerged) biofilms under well-defined flow conditions. 
Full details are given in Supplemental Material \cite{suppmat}; here we summarize the key
features. Flagella-less mutants of \textit{Bacillus subtilis} were used to avoid secondary
contributions to biofilm spreading \cite{Seminara12}.  Cells in exponential growth phase
were centrifuged and resuspended in growth medium before being loaded at the centre of 
Y04-D plates linked to the CellASIC ONIX microfluidic platform (EMD Millipore), and kept at 
$30$ $^\circ$C. In this setup, they are confined between glass and an overlying PDMS
sheet of thickness $d=114 \, \mu$m, with an initial gap of $h=6$ $\mu$m. 
Fresh medium was flowed through the chamber with a mean speed of 
$\sim 16 \, \mu$ms$^{-1}$ \cite{Martinez-Corral19,Liu15,Humphries17}.
Biofilm growth was imaged at $1$ frame/minute on a spinning-disc confocal microscope in 
bright field.  As the biofilms were often frilly, with long thin strands of 
matrix polymer protruding from their edges, a Gaussian image processing filter in MATLAB
was used to neglect these strands when identifying the interface with a Sobel edge detector.

Figure \ref{fig:3}(a) is a montage of the expanding biofilm edge and the best-fit circle for one particular 
experiment, while Figure \ref{fig:3}(b) plots the scaled biofilm radius $\mathcal{R}$ as a function of time. In a clear 
departure from unconfined bacterial biofilms, the ${\cal R}$ initially grows as a power law before tending
to saturate at long times. These profiles exhibit the main qualitative features predicted by the theoretical 
model for $\Xi > 1$. The lines of best fit (black lines in \ref{fig:3}(b), \cite{suppmat}) show good agreement 
over the entire time course of the experiments. A further comparison with theory is obtained by 
measuring in three different experiments, at the same nutrient concentration, the radius $\mathcal{R}(t_0)$ 
at a particular time $t_0 = 5\,$h, chosen
as a time when the biofilm radius had a least doubled from its initial value. The parameter $g$ relating
absolute and rescaled times was fitted across all experiments, and gives the value $\tau_0=4.29$ used in
Fig. \ref{fig:2}(b), while $\Xi$ is fitted independently for each.
These experimental points in the $\Xi-{\cal R}$ plane  are shown as blue circles in Figure \ref{fig:2}b), and agree 
very well with the poroelastic model developed here. 

Motivated by the desire to understand the evolution of biofilms under confinement, we have 
constructed a minimal mathematical model that uses a poro-elastic framework. This admits a family of 
self-similar quasi-steady solutions, parameterized by a dimensionless parameter $\Xi$ that measures
the elasticity of the matrix.  Those solutions are consistent with the experimentally observed behavior
of confined {\it B. subtilis} biofilms. For comparison, \cite{suppmat} presents the corresponding theoretical 
model in which, following previous work in the literature, the biomass is modelled instead 
as a viscous Newtonian fluid, neglecting the intrinsic elasticity of the biofilm ECM. 
In that case, a solution with power law growth tending to a maximum finite biofilm radius is not supported, 
demonstrating that modelling the matrix elasticity is essential to capturing biofilm growth under elastic confinement. 

Unlike unconfined biofilms, a subset of these solutions (where $\Xi > 1$) have a maximum radius due to 
a balance between elastic stresses and the osmotic pressure difference across the interface. The key parameter 
that determines which regime the system lies in and thus whether the biofilm grows predominately radially or 
axially is the stiffness of the biofilm matrix. Hence, we may view matrix elasticity is a competitive trait that 
could well be optimized by natural selection. 

\begin{acknowledgments}
We are grateful to G.G. Peng and J.A. Neufeld for discussions and P.A. Haas and A. Chamolly for valuable comments 
on an earlier version of the manuscript.  This work was supported in part by the Engineering and Physical 
Sciences Research Council, through a Doctoral Training Fellowship (GTF) and an Established Career Fellowship 
EP/M017982/1 (REG), and by a Wellcome Trust Interdisciplinary Fellowship and Discovery Fellowship 
BB/T009098/1 from the Biotechnology and Biological Sciences Research Council (NMO).
\end{acknowledgments}


\setcounter{equation}{0}
\setcounter{figure}{0}
\setcounter{table}{0}
\setcounter{page}{1}
\makeatletter
\renewcommand{\theequation}{S\arabic{equation}}
\renewcommand{\thefigure}{S\arabic{figure}}
\renewcommand{\bibnumfmt}[1]{[S#1]}
\renewcommand{\citenumfont}[1]{S#1}

\newpage

\large{\bf Supplementary Material}

\medskip

    We present dimensionless shallow layer scalings that reduce the system of equations describing 
    the full system (2)-(6) in the main text to a pair of coupled differential equations for the 
    height $h(r,t)$ and vertically averaged biomass volume fraction $\langle \phi \rangle(r,t)$ as a 
    function of radial distance $r$ and time $t$. The deformation $\boldsymbol{\xi}$ 
    is expressed as a function of derivatives of $h$, utilizing both global biomass volume conservation and a 
    pressure condition at the biofilm interface. The system is closed with boundary conditions for $h$ at the biofilm interface,
    obtained by extending the  framework outside the biofilm to the whole domain and imposing far-field 
    free-beam and zero-pressure conditions.

\section{Experimental}

\subsection{Supplementary information on methods and materials}
All experiments reported here used flagella-null cells of \textit{Bacillus subtilis} (NCIB 3610 hag::tet, a gift from 
Roberto Kolter).  Flagellaless cells were preferred because their inability to swim largely avoids contamination 
of inlets loaded with fresh growth medium in the microfluidic devices, and removes motility as a secondary 
contribution to biofilm spreading, as in earlier work \cite{Seminara12}. 

For each experiment, \textit{Bacillus subtilis} cells were streaked from $-80^\circ$C freezer stocks onto $1.5\%$ agar 
LB plates and incubated at $37^\circ$C for $12$ hours. Cells from a single colony were then inoculated in 
LB Broth (Lennox) at $37^\circ$C for $3$ hours to obtain cells in the exponential growth phase. These were 
centrifuged at $2600$ rpm for $6$ minutes and re-suspended with fresh minimal salts glycerol glutamate (MSgg), 
the standard biofilm growth medium for \textit{B. subtilis} \cite{Seminara12}. This MSgg medium contained $5$ mM 
potassium phosphate buffer (pH $7.0$), $100\,$mM MOPS buffer (pH $7.0$), $2\,$mM MgCl$_{2}$, $700 \mu\,$M CaCl$_{2}$, 
$50 \mu\,$M MnCl$_{2}$, $100\,\mu$M FeCl$_{3}$, $1 \, \mu$M ZnCl$_{2}$, $2 \, \mu$M thiamine HCl, $0.5\%$ (v/v) glycerol 
and $0.5\%$ (w/v) monosodium glutamate. 

Cells were then loaded at the center of Y04-D plates linked to the CellASIC ONIX microfluidic platform (EMD Millipore), 
and were incubated at $30$ $^\circ$C for the duration of each experiment. In this setup, they were confined between 
a rigid surface (glass) and an elastic sheet (PDMS, $114 \, \mu$m thick), a distance $6$ $\mu$m apart. In all experiments 
we flowed fresh MSgg medium via one inlet, using a pump pressure of $1$ psi, corresponding to a mean flow rate 
of ~$16 \, \mu$ms$^{-1}$ in the growth chamber \cite{Martinez-Corral19,Liu15,Humphries17}.
The subsequent growth of these submerged biofilms was then followed over time with a Zeiss Axio Observer Z1 microscope, 
connected to a Yokogawa Spinning Disk Confocal CSU and controlled by Zen Blue software. A Zeiss $10\times/0.3$ M27 
Plan-Apochromat objective lens was used to acquire bright-field images at a rate of 1 frame per minute. 
These images were analyzed using both the open source image processing package Fiji \cite{Schindelin12} and several 
custom MATLAB scripts utilizing MATLAB's Image Processing Toolbox. In particular, a Sobel edge detector was used to 
locate the biofilm edge. The experimental biofilms  were often frilly with long thin strands of matrix polymer protruding 
from the biofilm edge.  Hence, 2D gaussian filtering using the MATLAB inbuilt function imgaussfilt was used to neglect 
these strands when identifying where the interface is. In order to fit a circle to the extracted interface a least-squares 
fit was implemented.

\subsection{Raw experimental data}
\begin{figure}[t]
	\centering
	\includegraphics[clip, width=\columnwidth]{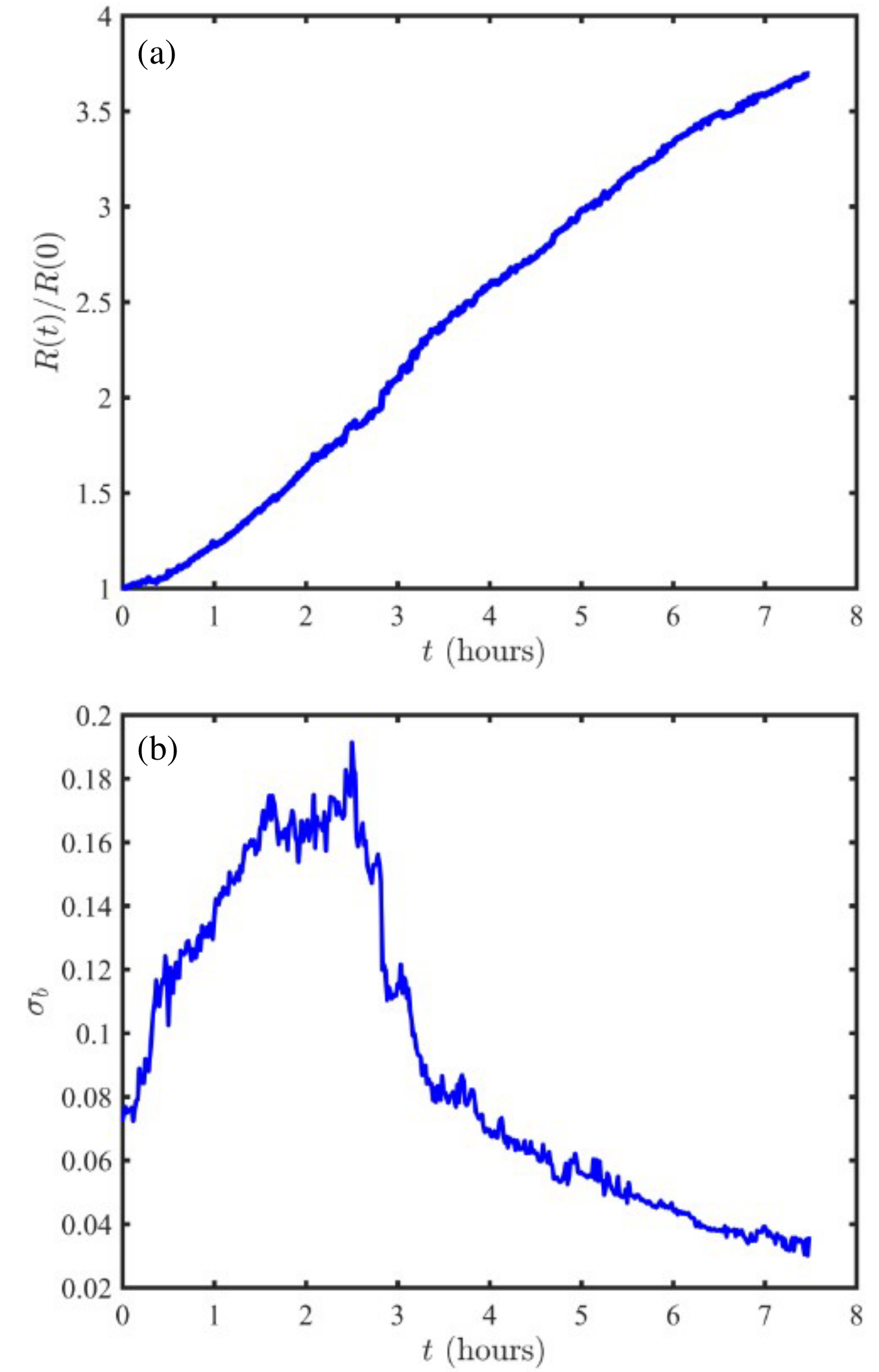}
	\caption{Raw data showing for a particular experiment how the scaled biofilm radius $\mathcal{R}$ (a) and the relative 
	deviation of the biofilm interface from a least-squares fitted circle $\sigma_b$ (b) vary as functions of time.
	}
	\label{fig:1suppmat}
\end{figure}

Figure \ref{fig:1suppmat} gives the corresponding raw data for the experiment given in the montage plot of 
Figure 3(a) of the main text, showing how in Figure \ref{fig:1suppmat}(a) the scaled biofilm radius $\mathcal{R}$ and in 
Figure \ref{fig:1suppmat}(b)  $\sigma_b$, a measure of the circularity of the biofilm edge, vary with time, 
where $\sigma_b$ satisfies
\begin{equation}
    \sigma_b = \text{std}\left( \boldsymbol{r_b} - \mathcal{R} \right) / \mathcal{R}.
\end{equation}
Here, $\boldsymbol{r_b}$ is a vector giving the scaled distance of the points on the biofilm edge from the center 
of the biofilm. As a biofilm grows, it becomes more circular (after an initial increase due to growth around an obstacle 
$\sigma_b$ decreases monotonically) but with frillier edges. Furthermore, as shown in the montage plot, 
interference fringes (Newton rings) are used to gain a qualitative understanding of how the upper PDMS sheet deforms. 
In particular, the fringes are circular, implying that the sheet deforms asymmetrically and thus evolves consistently with 
one of the key assumptions of the theoretical model, namely that $h = h(r,t)$ is independent of $\theta$.

\subsection{Fitting Procedure} \label{fittingprocedure}
Numerical solutions of (\ref{eq:Rgoverningequation}) predict the evolution of $\mathcal{R}$ as a function of 
dimensionless time $\tau$, with a single fitted parameter $\Xi$. To convert back to real time, the biofilm growth 
timescale $\tau_0 = g^{-1}$ has to be determined. This was found through an iterative procedure, utilising all three 
experimental datasets to obtain a series of increasingly accurate estimates for $g$, $\{ g_1, \, g_2, \, \cdots \}$, 
using the recursion relation that $g_{n+1}$ is the $g$ that minimises
\begin{equation}
\sum_i \left\{\underset{j}{\mathrm{avg}}\left(\left[\mathcal{R}_e(t_j)-\mathcal{R}_{\Xi_i}\left( g t_j \right) \right]^2 \right)\right\}_i,     
\end{equation}
where $i$ iterates over all datasets and $j$ over all points within the experimental dataset 
$\mathcal{R}_e(t_j) = R_e(t_j)/R_e(0)$ enumerated by $i$. $\mathcal{R}_{\tilde{\Xi}}(\tilde{\tau})$ is the solution 
to (\ref{eq:Rgoverningequation}) that is numerically computed for $\tau = \tilde{\tau}$ and $\Xi = \tilde{\Xi}$. 
$\Xi_i$ is the value of $\Xi$ that for the data set enumerated by $i$ minimises the objective function
\begin{equation}
    \underset{j}{\mathrm{avg}}\left(\left[\mathcal{R}_e(t_j)-\mathcal{R}_{\Xi}\left( g_n t_j \right) \right]^2 \right).  
\end{equation}
Here, all minimizations were performed using the MATLAB inbuilt function fminbnd \cite{Brent73}. This resulted in fitted 
values for the biofilm growth time scale of $g = 0.8574$ and for $\Xi$ of $1.7352$, $1.6702$ and $1.3358$ for the 
three different experiments.

\section{Full Poroelastic Framework}
Below, we denote the region which the biofilm occupies ($r \leq R$) the inner region and the region outside of the biofilm
($r \geq R$) the outer region.

\subsection{Inner Dimensional Vertical Boundary Conditions}
Since horizontal motion of the upper PDMS sheet can be neglected, imposing no-slip boundary conditions at both the 
lower and upper boundaries yields
\begin{subequations}
\begin{equation}
    w_f = w_s = u_s = \zeta = \xi = 0 \text{ at } z = 0, \label{eq:bc1a}
\end{equation}
\begin{equation}
    u_s = u_f = 0 \, , \, w_f = w_s = \frac{\partial H}{\partial t} \text{ at } z = H. \label{eq:bc1b}
\end{equation} \label{eq:bc1}
\end{subequations}
Vertically integrating (3a) using these boundary conditions and (1) gives the continuity equation for vertically 
averaged biomass
\begin{equation}
    \frac{\partial}{\partial t} \left( h \langle \phi \rangle \right) + \frac{1}{r}\frac{\partial}{\partial r}
    \left( r h \langle \phi \rangle \langle u_s \rangle \right) = g h \langle \phi \rangle. \label{eq:governcontinuity}
\end{equation}
Applying global biomass conservation, the biomass volume $V = 2\pi\int^R_0 r h\langle\phi\rangle dr$ satisfies 
$\partial V/\partial t = gV$. Expanding this out using the continuity equation (\ref{eq:governcontinuity}), (1) 
and the boundary conditions in (\ref{eq:bc1}) gives
\begin{equation}
    \frac{\partial R}{\partial t} = \frac{\langle \phi u_s \rangle}{\langle \phi \rangle}\Big{|}_{R} = \langle u_s \rangle \Big{|}_{R}. \label{eq:bc2}
\end{equation}
Modelling the upper sheet as a thin elastic beam, the pressure difference across the sheet is
\begin{equation}
[\Delta \tilde{p}]^+_- = B \nabla^4 h - \gamma \nabla^2 h. \label{eq:bc3}
\end{equation}
Balancing normal stress at the interface between the biofilm and the sheet yields
\begin{eqnarray}
	B\nabla^4 h - \gamma \nabla^2 h &=& \tilde{p} \big{|}_{h} - \left(K + 4G/3 \right)\frac{\partial \zeta}{\partial z} 
	\Big{|}_{h} \nonumber \\
	&-& (K - 2G/3)\frac{1}{r}\frac{\partial}{\partial r}(r \xi) \Big{|}_{h}\Longrightarrow \nonumber
\end{eqnarray}
\begin{eqnarray}
    p \big{|}_{h} &=& B\nabla^4 h - \gamma \nabla^2 h + \left(K + 4G/3 \right)\frac{\partial \zeta}{\partial z} \Big{|}_{h} \nonumber \\
    &+& \frac{K - 2G/3)}{r}\frac{\partial}{\partial r}(r \xi) \Big{|}_{h} - \frac{\Pi_{\text{os}} \phi^3}{(1-\phi)^3} \Bigg{|}_{h}, \label{eq:bc4}
\end{eqnarray}
where $\Pi_{\text{os}} = k_B T \beta^3 / 3\nu_0$.

\subsection{Dimensionless shallow-layer scalings}
We scale radial and vertical lengths with the initial radius $R_0 = R(t = 0)$ and height $H_0 = h(r = 0, t = 0)$ of 
the biofilm respectively i.e $\{ r, \, R \} \sim R_0$ and $\{ z, \, h \} \sim H_0$. Since the characteristic time scale 
for the system is that for biofilm growth, we scale $t \sim 1/g$. Utilizing 3) and (\ref{eq:bc2}), we 
find $\{ u_f, \, u_s \} \sim U_s = g R_0$ and $\{ w_f, \, w_s \} \sim g H_0$. Since $\boldsymbol{u_s}$ is defined 
as the material derivative of $\boldsymbol{\xi}$, we have $\xi \sim R_0$ and $\zeta \sim H_0$. By definition 
$\kappa \sim \kappa_0$. Finally, a leading order contribution to the pressure comes from the vertical confinement, 
i.e. (\ref{eq:bc4}) implies $p \sim P_0 = B H_0 / R_0^4$. We denote the dimensionless form of a function $f$ by $f^{*}$ and set for clarity
\begin{equation}
    \rho = r^{*} = \frac{r}{R(0)}, \ {\cal H} = h^{*} = \frac{h(r,t)}{h(0,0)},\nonumber
\end{equation}
\begin{equation}
    \tau = \tau^{*} = g t, \ {\mathcal{R}} = R^{*} = \frac{R(t)}{R(0)}, \ \mathcal{P} = p^{*} = \frac{p}{P_0}. \nonumber
\end{equation}
\subsection{Inner Governing Equations}
Using these scalings and setting $\epsilon = H_0/R_0$, the system of equations (3)-(9) becomes
\begin{subequations}
\begin{equation}
    \frac{\partial \phi}{\partial \tau} + \frac{\partial}{\partial z^{*}}\left( \phi w_s^{*} \right) + \frac{1}{\rho}\frac{\partial}{\partial \rho}\left( \rho \phi u_s^{*} \right) = \phi, \label{eq:nondimengovern1}
\end{equation}
\begin{eqnarray}
    -\frac{\partial \phi}{\partial t} &+& \frac{\partial}{\partial z^{*}}\left( (1 - \phi)w^{*}_f \right) \nonumber \\
    &+& \frac{1}{\rho}\frac{\partial}{\partial \rho}\left( \rho (1 - \phi) u^{*}_{f} \right) = -\phi, \label{eq:nondimengovern2}
\end{eqnarray}
\begin{equation}
    u_s = \frac{\rho}{\mathcal{R}}\frac{\partial \mathcal{R}}{\partial \tau}\frac{6z^{*}(\mathcal{H} - z^{*})}{\mathcal{H}^2}, \label{eq:nondimengovern3}
\end{equation}
\begin{equation}
    W_{1} \left(w^{*}_f - w^{*}_s \right) = -\frac{\kappa^{*}}{(1 - \phi)} \frac{\partial \mathcal{P}}{\partial z^{*}}, \label{eq:nondimengovern4}
\end{equation}
\begin{equation}
 \frac{\partial \mathcal{P}}{\partial \rho} =  \chi\frac{\partial^2 \xi^{*}}{\partial {z^{*}}^2} + \mathcal{O}\left(\chi \epsilon^2 \right), \label{eq:nondimengovern5}
\end{equation}
\begin{eqnarray}
\left(\frac{1}{W_1}\frac{\partial \mathcal{P}}{\partial z^{*}}\right) = P_1\Bigg{(} \frac{\partial^2 \zeta^{*}}{\partial {z^{*}}^2} &+& \frac{\tilde{K}}{\rho}\frac{\partial}{\partial \rho}\left( \rho \frac{\partial \xi^{*}}{\partial z^{*}} \right) \nonumber \\ 
&+& \mathcal{O}\left( \epsilon^2 \right) \Bigg{)}, \label{eq:nondimengovern6}
\end{eqnarray} \label{eq:nondimengovern}
\end{subequations}
\noindent together with the continuity equation for vertically averaged biomass
\begin{equation}
    \frac{\partial}{\partial \tau}\left( \mathcal{H}\varphi \right) + \frac{1}{\rho}\frac{\partial}{\partial \rho}\left( \rho \mathcal{H}\varphi v_s  \right) = \mathcal{H}\varphi, \label{eq:nondimengovern7}
\end{equation}
and corresponding vertical boundary conditions
\begin{subequations}
\begin{equation}
    w^{*}_f = w^{*}_s = u^{*}_s = \zeta^{*} = \xi^{*} = 0 \text{ at } z^* = 0, \label{eq:nondimenbc1}
\end{equation}
\begin{equation}
    u^{*}_s = u^{*}_f = 0 \, , \, w^{*}_f = w^{*}_s = \frac{\partial {\cal H}}{\partial t} \text{ at } z^* = {\cal H}. \label{eq:nondimenbc2}
\end{equation}
\begin{align}
    &\mathcal{P}\big{|}_{z^* = \mathcal{H}} = \nabla^4 \mathcal{H} - \gamma^{*}\nabla^2 \mathcal{H} - \Pi_{\text{os}}^{*} \frac{\phi^3}{(1 - \phi)^3}\Bigg{|}_{\mathcal{H}} \nonumber \\
    &+ P_s \Bigg{(} \frac{\partial \zeta^{*}}{\partial z^{*}}\Big{|}_{\mathcal{H}} + \frac{1}{\rho}\left(\frac{K - 2G/3}{K + 4G/3}\right)\frac{\partial}{\partial \rho}\left( \rho \xi^{*} \right)\Big{|}_{\mathcal{H}} \Bigg{)}, \label{eq:nondimenbc3}
\end{align} \label{eq:nondimenbc}
\end{subequations}
\noindent where $\boldsymbol{u_s^{*}} = (u_s, \, w_s)$ can be expressed as the material derivative of $\boldsymbol{\xi^{*}} = (\xi, \, \zeta)$ using
\begin{subequations}
\begin{equation}
    w_s^{*} = \frac{\partial \zeta}{\partial \tau} + w_s^{*} \frac{\partial \zeta^{*}}{\partial z^{*}} + u_s^{*} \frac{\partial \zeta^{*}}{\partial \rho}, \label{eq:nondimengovern8}
\end{equation}
\begin{equation}
    u_s^{*} = \frac{\partial \xi^{*}}{\partial \tau} + w_s^{*} \frac{\partial \xi^{*}}{\partial z^{*}} + u_s^{*} \frac{\partial \xi^{*}}{\partial \rho}, \label{eq:nondimengovern9}
\end{equation}
\end{subequations}
\noindent while $\{ \tilde{K}, \, W_1, \, \chi, \, P_1, \, \Pi_{\text{os}}^{*}, \, \gamma^{*}, \, P_s \}$ are non-dimensional constants that satisfy
\begin{subequations}
\begin{equation}
    \tilde{K} = \frac{K + G/3}{K + 4G/3}, \quad W_1 = \frac{\mu_f g H_0^2}{\kappa_0 P_0}, \quad \chi = \frac{G}{\epsilon^2 P_0},
\end{equation}    
\begin{equation}    
    \quad P_1 = \frac{\kappa_0 \left( K + 4G/3 \right)}{\mu_f g H_0^2}, \, \Pi_{\text{os}}^{*} = \frac{R_0^4 \, \Pi_{\text{os}}}{B H_0},
\end{equation}    
\begin{equation}    
    \gamma^{*} = \frac{\gamma R_0^2}{B}, \, P_s = \frac{R_0^4 \left( K + 4G/3 \right)}{B H_0}.
\end{equation} \label{eq:nondimensionalconstants}
\end{subequations}
\noindent Here, $\{ W_1, \, \chi, \, P_1 \}$ are dimensionless measures of the ability of flow to generate a vertical pressure gradient and the relative strength of the pressure gradients compared to elastic stresses in the horizontal and vertical respectively. $\gamma^{*}$ and $\Pi_{\text{os}}^{*}$ are the non-dimensional surface tension and osmotic pressure scaling groups. Finally, 
$P_s$ measures the relative strength of the elastic stresses from the biofilm and the PDMS sheet at the upper interface. 

\subsection{Order of Magnitude Estimates for Parameters} \label{orderofmagnitudeestimates}
In a typical experiment, the biofilm initially has height $H_0 \sim 10^{-5}\si{\metre}$ and radius $R_0 \sim 10^{-4}\si{\metre}$. 
We assume that the dynamic viscosity of the nutrient rich liquid phase can be approximated by that of water, 
$\mu_f \sim 7.98 \times 10^{-4}\si{\pascal\second }$. From Seminara $\etal$, an order of magnitude estimate for the biofilm growth 
rate $g$ is $g^{-1} \sim \SI{2.3}{\hour}$ \cite{Seminara12}. Furthermore, the characteristic biofilm permeability scale $\kappa_0 \sim \xi^2_0$ where the biofilm mesh length scale $\xi_{\infty} \sim \SI{50}{\nano\metre}$ i.e. 
$\kappa_0 \sim 2.5 \times 10^{-15}\si{\metre^2}$

Picioreanu $\etal$ estimated the mechanical properties of a range of different biofilms cultivated from activated 
sludge supernatant using optical coherence tomography, obtaining an effective Poisson ratio $\nu_b = 0.4$ and 
Young's modulus in the range $70 - \SI{700}{\Pa}$ \cite{Picioreanu18}. Assuming isotropy, we can thus estimate $K$ and $G$ 
as being in the range $K = E_b/3(1 - 2\nu_b) \sim 117 - \SI{1170}{\Pa}$ and $G = E_b/2(1 + 2 \nu_b) 
\sim 19.4 - \SI{194}{\Pa}$ respectively. 

The PDMS sheet has thickness $d \sim 10^{-4}\si{\metre}$, Poisson's ratio $\nu \sim 0.5$ and Young's modulus 
$E \sim 1.9 \times 10^{6} \, \si{\pascal}$ (a value of $55$ measured using a type A durometer). The matrix solid fraction $\beta$ 
and the volume occupied by one monomer of extracellular matrix varies considerably, depending on a range of factors 
such as the species of bacteria and the nutrient concentration. Aiming to show that the osmotic pressure contribution 
can be neglected, we consider uppper and lower bounds for $\beta$ and $\nu_0$ respectively i.e. 
$\beta = \mathcal{O}(1)$ and $\nu_0 \sim 10^{-24} \si{\metre^3}$.  Finally, we estimate the surface tension 
between the biofilm and the sheet using that between water and PDMS 
($\gamma \sim 4\times 10^{-2}\si{\newton\metre^{-2}}$) \cite{Ismail09}.

Hence, estimating values for the non-dimensional parameters $\{ \epsilon, \, W_1, \, \chi, \, P_1, \, \gamma^{*}, \, \Pi_{\text{os}}^{*}, \, P_{s}, \, \epsilon_{\text{outer}}  \}$ gives
\begin{subequations}
\begin{equation}
    \epsilon = \frac{H_0}{R_0} \sim 10^{-1} \ll 1, \label{eq:epsilonestimate}
\end{equation}
\begin{eqnarray}
    W_1 &=& \frac{12\mu_f g H_0 R_0 (1 - \nu^2)}{\kappa_0 E} \nonumber \\
    &\sim& 1.83 \times 10^{-7} \ll 1, \label{eq:W1estimate}
\end{eqnarray}
\begin{eqnarray}
    \chi &=& \frac{12 G (1 - \nu^2)}{\epsilon^3 E}\left( \frac{R_0}{d} \right)^3 \nonumber \\
    &\sim& 9.19\times\{ 10^{-2} - 10^{-1} \}, \label{eq:chiestimate}
\end{eqnarray}
\begin{equation}
    P_1 = \frac{\kappa_0 (K + 4G/3)}{\mu_f g H_0^2} \sim 3.71 \times \{ 10^{4} - 10^{5} \} \gg 1, \label{eq:P1estimate}
\end{equation}
\begin{equation}
    \gamma^{*} = \frac{12 \gamma (1 - \nu^2)}{E d}\left( \frac{R_0}{d} \right)^2 \sim 2.25 \times 10^{-3} \ll 1, \label{eq:gammaestimate}
\end{equation}
\begin{eqnarray}
    \Pi_{\text{os}} &=& \frac{4 k_{B} T (1 - \nu^2) \beta^3 }{ \epsilon \nu_0 E }\left( \frac{R_0}{d} \right)^3 \nonumber \\
    &\sim& 6.60 \times 10^{-2} \ll 1, \label{eq:Piosestimate}
\end{eqnarray}
\begin{eqnarray}
    P_s &=& \frac{12\left( 1 - \nu^2 \right) \left( K + 4G/3 \right)}{\epsilon E}\left( \frac{R_0}{d} \right)^3 \nonumber \\
    &\sim& 6.77 \times \{ 10^{-3} - 10^{-2} \} \ll 1, \label{eq:Psestimate}  
\end{eqnarray}
\begin{eqnarray}
    \epsilon_{\text{outer}} &=& \frac{144 g \mu_f \left( 1 - \nu^2 \right)}{\epsilon^3 E}\left( \frac{R_0}{d} \right)^3 \nonumber \\
    &\sim& 5.48 \times 10^{-9} \ll 1. \label{eq:epsilonouterestimate}
\end{eqnarray}
\end{subequations}
\subsection{Stiff Elastic Confinement}
Hence, under experimental conditions, we see that the upper elastic sheet is sufficiently stiff that 
$\{ W_1, \, \gamma^{*}, \, \Pi^{*}_{\text{os}}, \, P_s \} \ll 1$ i.e. the dominant contribution to the pressure 
arises from the upper confinement. $P_1 \gg 1$ means that elastic stresses dominate the vertical pressure gradient. 
In general, $\chi = \mathcal{O}(1)$. Hence, the systems of governing equations given in $(\ref{eq:nondimengovern}) - (\ref{eq:nondimenbc})$ reduces to 
\begin{subequations}
\begin{equation}
    \frac{\partial^2 \xi^{*}}{\partial {z^{*}}^2} = \frac{1}{\chi}\frac{\partial \mathcal{P}}{\partial \rho}, \label{eq:stiffgoverning1}
\end{equation}
\begin{equation}
\frac{\partial^2 \zeta^{*}}{\partial {z^{*}}^2} = -\frac{\tilde{K}}{\rho}\frac{\partial}{\partial \rho}\left( \rho \frac{\partial \xi^{*}}{\partial z^{*}} \right),  \label{eq:stiffgoverning2}  
\end{equation}
\begin{equation}
    \frac{6\rho}{\mathcal{R}}\frac{\partial \mathcal{R}}{\partial \tau} \frac{z^{*}(\mathcal{H} - z^{*})}{{\mathcal{H}}^2}\left( \left( 1 - \frac{\partial \zeta^{*}}{\partial z^{*}} \right)\left( 1 - \frac{\partial \xi^{*}}{\partial \rho} \right) - \frac{\partial \xi^{*}}{\partial z^{*}}\frac{\partial \zeta^{*}}{\partial \rho} \right) \nonumber 
\end{equation}    
\begin{equation}
    = \frac{\partial \xi^{*} }{\partial z^{*}}\frac{\partial \zeta^{*}}{\partial \tau} + \frac{\partial \xi^{*}}{\partial \tau}\left( 1 - \frac{\partial \zeta^{*}}{\partial z^{*}} \right), \label{eq:stiffgoverning3}
\end{equation}
\end{subequations}
with corresponding boundary conditions 
\begin{subequations}
\begin{equation}
    \left[\frac{\partial \xi^{*}}{\partial \tau} \right]_{z^{*} = 0} = 0, \label{eq:stiffbc1}
\end{equation}    
\begin{equation}    
    \left[ \frac{\partial \xi^{*}}{\partial \tau} + \frac{\partial \mathcal{H}}{\partial \tau}\frac{\partial \xi^{*}}{\partial z^{*}} \right]_{z^{*} = \mathcal{H}} = 0, \label{eq:stiffbc2}
\end{equation}
\begin{equation}
    \left[\frac{\partial \zeta^{*}}{\partial \tau} \right]_{z^{*} = 0} = 0, \label{eq:stiffbc3}
\end{equation}    
\begin{equation}    
    \left[ \frac{\partial \zeta^{*}}{\partial \tau} + \frac{\partial \mathcal{H}}{\partial \tau}\frac{\partial \zeta^{*}}{\partial z^{*}} \right]_{z^{*} = \mathcal{H}} = \frac{\partial \mathcal{H}}{\partial \tau}. \label{eq:stiffbc4}
\end{equation}
\end{subequations}
where
\begin{equation}
    \mathcal{P} = \nabla^4 \mathcal{H}. \label{eq:stiffpressure}
\end{equation}
Since $\mathcal{P}$ is independent of $z^{*}$, integrating (\ref{eq:stiffgoverning1}) twice with respect to 
$z^{*}$ yields the functional form for $\xi^{*}$ 
\begin{equation}
\xi = b_0 + b_1 z^{*} + b_2 {z^{*}}^2, \label{eq:xifunctionalform}
\end{equation}
where $\{ b_i = b_i(\rho,\tau) : i \in [ 0, \, 1, \, 2 ] \}$ are independent of $z^{*}$ and $b_2$ satisfies
\begin{subequations}
\begin{equation}
    b_2 = \frac{1}{2\chi} \frac{\partial \mathcal{P}}{\partial \rho} = \frac{1}{2 \chi} \frac{\partial}{\partial \rho}\left( \nabla^4 \mathcal{H} \right). \label{eq:defb2}
\end{equation}
(\ref{eq:stiffbc1}) and (\ref{eq:stiffbc2}) simplify respectively to give
\begin{equation}
    \frac{\partial b_0}{\partial \tau} = 0 \Longrightarrow b_0 = b_0(\rho). \label{eq:defb0}
\end{equation}
\begin{equation}
    \frac{\partial}{\partial \tau}\left( \mathcal{H} b_1 + \mathcal{H}^2 b_2 \right) = 0 \Longrightarrow b_1 = \frac{B}{\mathcal{H}} - \mathcal{H}b_2, \label{eq:defb1}
\end{equation}
\end{subequations}
where $B = B(\rho)$ is independent of $\tau$ and $z^{*}$. Hence, (\ref{eq:stiffgoverning2}) becomes
\begin{eqnarray}
    \frac{\partial^2 \zeta^{*}}{\partial {z^{*}}^2} &=& -\frac{\tilde{K}}{\rho}\frac{\partial}{\partial \rho}\left( \rho \frac{\partial \xi^{*}}{\partial z^{*}} \right) \nonumber \\
    &=&  -\frac{\tilde{K}}{\rho}\frac{\partial}{\partial \rho}\left( \frac{\rho B}{\mathcal{H}} + \rho b_2 (2z^{*} - \mathcal{H}) \right)
\end{eqnarray}
Integrating this twice with respect to $z^{*}$ yields the functional form for $\zeta^*$,
\begin{equation}
        \zeta^{*} = a_0 + a_1 z^{*} + a_2 {z^{*}}^2 + a_3 {z^{*}}^3, \label{eq:zetafunctionalform}
\end{equation}
where $\{ a_i = a_i(\rho,\tau) : i \in [ 0, \, 1, \, 2, \, 3 ] \}$ are independent of $z^{*}$ and $a_2$ and $a_3$ satisfy
\begin{subequations}
\begin{equation}
    a_2 = \frac{\tilde{K}}{2\rho}\frac{\partial}{\partial \rho}
    \left( \rho \mathcal{H} b_2 - \frac{\rho B}{\mathcal{H}} \right), \label{eq:defa2}
\end{equation}
\begin{equation}
    a_3 = -\frac{\tilde{K}}{3\rho}\frac{\partial}{\partial \rho}\left( \rho b_2 \right). \label{eq:defa3}
\end{equation}
(\ref{eq:stiffbc3}) and (\ref{eq:stiffbc4}) simplify respectively to give
\begin{equation}
    \frac{\partial a_0}{\partial \tau} = 0 \Longrightarrow a_0 = a_0(\rho). \label{eq:defa0}
\end{equation}
\begin{equation}
    \frac{\partial}{\partial \tau}\left( \mathcal{H} a_1 + 
    \mathcal{H}^2 a_2 + \mathcal{H}^3 a_3 \right) = \frac{\partial \mathcal{H}}{\partial \tau} \Longrightarrow \nonumber
\end{equation}    
\begin{equation}    
    a_1 = 1 + \frac{A}{\mathcal{H}} - \mathcal{H}a_2 - \mathcal{H}^2a_3, \label{eq:defa1}
\end{equation}
where $A = A(\rho)$ is independent of $\tau$ and $z^{*}$.
\end{subequations}
Substituting (\ref{eq:xifunctionalform}) and (\ref{eq:zetafunctionalform}) into (\ref{eq:stiffgoverning3}) and 
equating the various powers of $z^{*}$ gives the following set of six coupled equations for the variables $a_j$ 
and $b_k$ where $j \in [0, \, 1, \, 2, \, 3]$ and $k \in [0, \, 1, \, 2]$
\begin{subequations}
\begin{equation}
    \frac{\partial b_1}{\partial \tau} + b_1 \frac{\partial a_1}{\partial \tau} - a_1 \frac{\partial b_1}{\partial \tau} = \nonumber
\end{equation}
\begin{equation}
    \frac{6\rho}{\mathcal{R}\mathcal{H}} \frac{\partial  \mathcal{R}}{\partial \tau} \left( \left(1 - a_1 \right)\left( 1 - \frac{\partial b_0}{\partial \rho}\right) - b_1 \frac{\partial a_0}{\partial \rho} \right),
\end{equation}
\begin{equation}
    (1 - a_1)\frac{\partial b_2}{\partial \tau} - 2 a_2 \frac{\partial b_1}{\partial \tau} 
    + 2 b_2 \frac{\partial a_1}{\partial \tau} + b_1 \frac{\partial a_2}{\partial \tau} \nonumber
\end{equation}
\begin{equation}
     = -\frac{6\rho}{\mathcal{R}\mathcal{H}^2} \frac{\partial  \mathcal{R}}{\partial \tau} 
     \left( \left(1 - a_1 \right)\left( 1 - \frac{\partial b_0}{\partial \rho}\right) 
     - b_1 \frac{\partial a_0}{\partial \rho} \right) \nonumber
\end{equation}
\begin{eqnarray}
    + \frac{6\rho}{\mathcal{R}\mathcal{H}} \frac{\partial  \mathcal{R}}{\partial \tau} \Bigg{(} &-&2a_2
    \left( 1 - \frac{\partial b_0}{\partial \rho}\right) - \frac{\partial b_1}{\partial \rho}\left( 1 - a_1 \right) \nonumber \\
    &-& 2b_2 \frac{\partial a_0}{\partial \rho} - b_1 \frac{\partial a_1}{\partial \rho} \Bigg{)}, 
\end{eqnarray}
\begin{equation}
    b_1\frac{\partial a_3}{\partial \tau} + 2 b_2 \frac{\partial a_2}{\partial \tau} - 2 a_2 \frac{\partial b_2}{\partial \tau} - 3 a_3 \frac{\partial b_1}{\partial \tau} \nonumber
\end{equation}
\begin{eqnarray}
     = -\frac{6\rho}{\mathcal{R}\mathcal{H}^2} \frac{\partial  \mathcal{R}}{\partial \tau} \Bigg{(} &-&2a_2\left( 1 - \frac{\partial b_0}{\partial \rho}\right) - \frac{\partial b_1}{\partial \rho}\left( 1 - a_1 \right) \nonumber \\
     &-& 2b_2 \frac{\partial a_0}{\partial \rho} - b_1 \frac{\partial a_1}{\partial \rho} \Bigg{)} \nonumber
\end{eqnarray}
\begin{eqnarray}
    + \frac{6\rho}{\mathcal{R}\mathcal{H}} \frac{\partial  \mathcal{R}}{\partial \tau} \Bigg{(} &-&3a_3\left( 1 - \frac{\partial b_0}{\partial \rho}\right) + 2 a_2 \frac{\partial b_1}{\partial \rho} - (1 - a_1)\frac{\partial b_2}{\partial \rho} \nonumber \\
    &-& b_1 \frac{\partial a_2}{\partial \rho} - 2b_2 \frac{\partial a_1}{\partial \rho} \Bigg{)}, 
\end{eqnarray}
\begin{equation}
    2 b_2 \frac{\partial a_3}{\partial \tau} - 3 a_3 \frac{\partial b_2}{\partial \tau} \nonumber
\end{equation}
\begin{eqnarray}
    = -\frac{6 \rho}{\mathcal{R}\mathcal{H}^2}\frac{\partial \mathcal{R}}{\partial \tau}\Bigg{(} &-& 3 a_3 \left( 1 - \frac{\partial b_0}{\partial \rho} \right) + 2 a_2 \frac{\partial b_1}{\partial \rho} \nonumber \\
    &-& (1 - a_1)\frac{\partial b_2}{\partial \rho} - b_1 \frac{\partial a_2}{\partial \rho} \nonumber \\
    &-& 2b_2 \frac{\partial a_1}{\partial r} \Bigg{)},
\end{eqnarray}
\begin{equation}
    2a_2 \frac{\partial b_2}{\partial \rho} + 3 a_3 \frac{\partial b_1}{\partial \rho} - b_1 \frac{\partial a_3}{\partial \rho} - 2 b_2 \frac{\partial a_2}{\partial \rho} = 0,
\end{equation}
\begin{equation}
    3 a_3 \frac{\partial b_2}{\partial \rho} - 2 b_2 \frac{\partial a_3}{\partial \rho} = 0.
\end{equation}
\end{subequations}
In particular, equating co-efficients of ${z^{*}}^6$ gives
\begin{equation}
    3 a_3 \frac{\partial b_2}{\partial \rho} = 2 b_2 \frac{\partial a_3}{\partial \rho}.
\end{equation}
We then have three possible cases:
\begin{enumerate}
\item Mode zero, $b_2 = 0 \Longrightarrow a_3 = 0$,
\item Mode one, $b_2 \neq 0$ but $a_3 = 0$,
\item Mode two, $b_2 \neq 0$ and $a_3 \neq 0$.
\end{enumerate}
\subsubsection{Mode zero}
When $b_2 = 0$, $\mathcal{H}$ satisfies the differential equation 
\begin{equation}
    \frac{\partial \mathcal{P}}{\partial \rho} = 0\frac{\partial}{\partial \rho}\left( \nabla^4 \mathcal{H} \right) = 0,
\end{equation}
which has the general solution
\begin{eqnarray}
    \mathcal{H} &=& A_{00} + A_{01} \, \rho^2 + A_{02} \, \rho^4 \nonumber \\
    &+& A_{03} \, \log{\rho} + A_{04} \, \rho^2 \log{\rho}.
\end{eqnarray}
Note that this mode is 
dominant in the limit $\chi \ll 1$.

\subsubsection{Mode one}
When $a_3 = 0$, $\mathcal{H}$ satisfies the differential equation
\begin{equation}
\frac{\partial}{\partial \rho}\left( r b_2 \right) = 0 \Longrightarrow b_2 = \frac{32}{\chi}\frac{A_{15}}{\rho} = \frac{1}{2\chi}\frac{\partial}{\partial \rho}\left( \nabla^4 \mathcal{H} \right) \Longrightarrow \nonumber
\end{equation}
\begin{equation}
\frac{\partial}{\partial \rho}\left( \nabla^4 \mathcal{H} \right) = \frac{64 A_{15} }{\rho},
\end{equation}
which has the general solution
\begin{eqnarray}
    \mathcal{H} &=& A_{10} + A_{11} \, \rho^2 + A_{12} \, \rho^4 + A_{13} \, \log{\rho} \nonumber \\
    &+& A_{14} \, \rho^2 \log{\rho} + A_{15} \, \rho^4 \log{\rho}.
\end{eqnarray}

\subsubsection{Mode two}
When both $b_2$ and $a_3$ $\neq 0$, $b_2$ satisfies the differential equation
\begin{equation}
\frac{\partial}{\partial \rho}\left( \frac{b_2^3}{a_3^2} \right) = 0 \Longrightarrow b_2^3 = -\frac{9 A_{25}}{2 \chi \rho^2} \left( \frac{\partial}{\partial \rho} \left(\rho b_2 \right) \right)^2.
\end{equation}
Employing the substitution $\tilde{b} = - \rho b_2$, this differential equation becomes separable and 
can be integrated to give
\begin{equation}
\int \tilde{b}^{3/2} d\tilde{b} = \int \sqrt{\frac{2 \chi}{9 \rho A_{25}}} \, d\rho \Longrightarrow 
\end{equation}
\begin{equation}
\tilde{b} = \frac{9 A_{25}}{2 \chi \rho} \Longrightarrow b_2 = -\frac{9 A_{25}}{2 \chi \rho^2}.
\end{equation}
Hence, $\mathcal{H}$ satisfies the differential equation
\begin{equation}
    \frac{\partial}{\partial \rho} \left( \nabla^4 \mathcal{H} \right) = - \frac{9 A_{25}}{\rho^2}, 
\end{equation}
which has the general solution
\begin{eqnarray}
    \mathcal{H} &=& A_{20} + A_{21} \, \rho^2 + A_{22} \, \rho^4 + A_{23} \, \log{\rho} \nonumber \\
    &+& A_{24} \, \rho^2 \log{\rho} + A_{25} \, \rho^3.
\end{eqnarray}

\subsection{Horizontal Boundary Conditions} 
\label{horizontalbc}
Define the inverse function of $\mathcal{R}(\tau)$, $\tau_1(\rho)$, as satisfying 
\begin{equation}
\tau_1(\rho) = \left\{
\begin{array} {ll}
\tau : \rho = \mathcal{R}(\tau) & \mbox{ when } \rho > \mathcal{R}(0) = 1,\\
0  & \mbox{otherwise}.  \\
\end{array}. 
\right. \\
\end{equation}
Hence, constraining the pressure of the solid phase $(\sigma^{*}_l)_{z^{*}z^{*}}$ to be constant at the biofilm interface yields
\begin{equation}
    \frac{\partial \zeta^{*}}{\partial z^{*}} + \frac{\tilde{K}}{\rho}\frac{\partial}{\partial \rho}
    \left( \rho \xi^{*} \right) = C_0 \Longrightarrow \nonumber \\
\end{equation}    
\begin{equation}    
    (a_1 - 1) + \frac{\tilde{K}}{\rho}\frac{\partial}{\partial \rho}\left( \rho b_0 \right) = C_0, \label{eq:pressurecondition1}
\end{equation}
at $\tau = \tau_1(\rho)$, where $C_0$ is a constant which is set from the initial pressure difference at $\tau = 0$ across the edge of the biofilm. From symmetry, ${\cal H}$ and ${\mathcal{P}}$ are even in $\rho$ at $\rho = 0$, i.e.
\begin{equation}
\frac{\partial {\cal H}}{\partial \rho}(0,\tau) = \frac{\partial {\mathcal{P}}}{\partial \rho}(0,\tau) = 0. \label{eq:centrebc}
\end{equation}
We assume that the biofilm grows uniformly at the interface, namely the vertically averaged biomass volume fraction $\varphi$ satisfies 
\begin{subequations}
\begin{equation}
    \varphi (\rho,\tau_1) = \varphi_{\infty}, \label{eq:phibc1}
\end{equation}    
\begin{equation}    
    \frac{\partial \varphi}{\partial \tau}(\rho,\tau_1) = 0, \label{eq:phibc2}
\end{equation}
\end{subequations}
where $\varphi_{\infty}$ is a constant. 

\subsection{Outer Governing equations}
 When $\rho > {\mathcal{R}}$, we have a lubrication flow of a single phase Newtonian fluid with viscosity $\mu_f$. 
 Hence, the vertically averaged fluid velocity $\langle u^{*} \rangle$ satisfies
\begin{equation}
    \langle u^{*} \rangle =-\frac{{\cal H}^2}{\epsilon_{\text{outer}}} \frac{\partial}{\partial \rho}\left( B \nabla^4 {\cal H} \right),
\end{equation}
leading to the continuity equation
\begin{equation}
    \epsilon_{\text{outer}} \frac{\partial {\cal H}}{\partial t} = \frac{1}{\rho}\frac{\partial}{\partial \rho}\left ( \rho {\cal H}^3 \frac{\partial}{\partial \rho}(B \nabla^4 {\cal H}) \right), \label{eq:fullcontinuityoutside}
\end{equation}
where the nondimensional constant $\epsilon_{\text{outer}}$ satisfies
\begin{equation}
    \epsilon_{\text{outer}} = \frac{12 g \mu_f}{\epsilon^2 P_0}.
\end{equation}
From above, under experimental conditions, $\epsilon_{\text{outer}} \ll 1$. Hence, for $\rho \ll \epsilon_{\text{outer}}^{-1/6}$ and $\{ \tau, \, \mathcal{H} \} = \mathcal{O}(1)$, (\ref{eq:fullcontinuityoutside}) becomes
\begin{equation}
\frac{1}{\rho}\frac{\partial}{\partial \rho}\left ( \rho {\cal H}^3 \frac{\partial}{\partial \rho}(B \nabla^4 {\cal H}) \right) = 0. \label{eq:reducedcontinuityoutside}    
\end{equation}

\subsection{Outer Boundary Conditions}
In practice, the PDMS sheet has a finite radial extent at $\rho = \mathcal{R}_{\text{outer}} = R_{\text{outer}}/R(0)$ where $1 \ll \mathcal{R}_{\text{outer}} \ll \epsilon_{\text{outer}}^{-1/6}$. There are two possible kinds of boundary conditions that could be imposed here.  If the sheet is clamped, namely fixed height and zero first derivative of height, we have
\begin{subequations}
\begin{equation}
    {\cal H}\left( \mathcal{R}_{\text{outer}},\tau  \right) = {\cal H}_{\infty},
\end{equation}
\begin{equation}
    \frac{\partial {\cal H}}{\partial \rho}\left(\mathcal{R}_{\text{outer}},\tau \right) = 0,
\end{equation} \label{eq:clampedbc}
\end{subequations}
\noindent where ${\cal H}_{\infty} = h_{\infty}/H_0$ is a constant. Alternatively, if the sheet is not clamped, we impose free-beam 
conditions at $\rho = \mathcal{R}_{\text{outer}}$, namely
\begin{subequations}
\begin{equation}
    \frac{\partial^2 {\cal H}}{\partial \rho^2}\left(\mathcal{R}_{\text{outer}},\tau \right) = 0.    
\end{equation}
\begin{equation}
    \frac{\partial^3 {\cal H}}{\partial \rho^3}\left(\mathcal{R}_{\text{outer}},\tau \right) = 0.
\end{equation}  \label{eq:notclampedbc}
\end{subequations}

\subsection{Interface Matching Conditions}
A fluid flux balance at the biofilm edge yields 
\begin{equation}
    \left( \langle (1 - \phi)u^{*}_f \rangle + \varphi \frac{\partial \mathcal{R}}{\partial \tau}  \right)\Bigg{|}_{\rho = \mathcal{R}^{-}}  = \langle u^{*} \rangle \Big{|}_{\rho = \mathcal{R}^{+}}. 
\end{equation}
At leading order in $\epsilon_{\text{outer}}$, this simplifies to 
\begin{equation}
\frac{\partial}{\partial \rho}\left( \nabla^4 {\cal H} \right)\Bigg{|}_{\rho = {\mathcal{R}}^{+}} = 0. \label{eq:fluidfluxmatchingcondition}
\end{equation}
This requires that across $\rho = \mathcal{R}$, fourth and lower derivatives of ${\cal H}$ are continuous.

\section{Simplification from a two to a one phase system}
In the above, we have written down a set of governing equations for the full two-phase system, 
considering both the inner and the outer regions, with corresponding boundary conditions at $\rho = 0, \mathcal{R}$ 
and $\mathcal{R}_{\text{outer}}$. Working in the limit that $\mathcal{R}_{\text{outer}} \gg 1$, here we simplify our 
framework to just considering a single phase system, namely the inner region, together with boundary conditions at 
$\rho = 0$ and $\mathcal{R}$. Utilising the general form of the solution for ${\cal H}$ in the outer region, we 
achieve this by re-writing the far field boundary conditions at $\rho = \mathcal{R}_{\text{outer}}$ (expressed in 
terms of derivatives of ${\cal H}$ at $\rho = \mathcal{R}_{\text{outer}}$) in terms of derivatives of ${\cal H}$ 
at $\rho = \mathcal{R}$, noting that these derivatives are continuous across the biofilm interface. 

\subsection{Matching Machinery}
Integrating (\ref{eq:reducedcontinuityoutside}), using the boundary condition given in (\ref{eq:fluidfluxmatchingcondition}), 
the general solution for ${\cal H}$ in the outer region is  
\begin{equation}
    {\cal H} = A_0 + A_1 \rho^2 + A_2 \rho^4 + A_3 \log{\rho} + A_4 \rho^2 \log{\rho}. \label{eq:outersolution}
\end{equation}
Defining vectors containing the constants of integration, the derivatives of ${\cal H}$ at the interface and the derivatives of ${\cal H}$ at the radial extent of the sheet, $\boldsymbol{A}_{\text{outer}}$, $\boldsymbol{{\cal H}}_{\text{interface}}$ and $\boldsymbol{{\cal H}}_{\text{outer}}$ respectively, as satisfying
\begin{subequations}
\begin{equation}
\boldsymbol{A}_{\text{outer}}= [A_0, \, A_1, \, A_2, \, A_3, \, A_4]^T,
\end{equation}
\begin{eqnarray}
 \boldsymbol{{\cal H}}_{\text{interface}} = [&&{\cal H}_0, \, {\cal H}_1, \, {\cal H}_2, \, {\cal H}_3, \, {\cal H}_4]^T \nonumber \\
 =\Big{[}&&{\cal H}(\mathcal{R},\tau), \, \frac{\partial {\cal H}}{\partial \rho}(\mathcal{R},\tau), \, \frac{\partial^2 {\cal H}}{\partial \rho^2}(\mathcal{R},\tau), \nonumber \\
&& \frac{\partial^3 {\cal H}}{\partial \rho^3}(\mathcal{R},\tau), \, \frac{\partial^4 {\cal H}}{\partial \rho^4}(\mathcal{R},\tau) \Big{]}^T,   
\end{eqnarray}
\begin{eqnarray}
 \boldsymbol{{\cal H}}_{\text{outer}} = \Big{[}&&{\cal H}(\mathcal{R}_{\text{outer}},\tau), \, \frac{\partial {\cal H}}{\partial \rho}(\mathcal{R}_{\text{outer}},\tau), \, \frac{\partial^2 {\cal H}}{\partial \rho^2}(\mathcal{R}_{\text{outer}},\tau), \nonumber \\
&& \frac{\partial^3 {\cal H}}{\partial \rho^3}(\mathcal{R}_{\text{outer}},\tau), \, \frac{\partial^4 {\cal H}}{\partial \rho^4}(\mathcal{R}_{\text{outer}},\tau) \Big{]}^T,   
\end{eqnarray}
\end{subequations}
we can express $\boldsymbol{{\cal H}}_{\text{outer}}$ in terms of $\boldsymbol{{\cal H}}_{\text{interface}}$ using (\ref{eq:outersolution}): 
\begin{equation}
\boldsymbol{H}_{\text{interface}} = \boldsymbol{\mathcal{M}}(\mathcal{R})\boldsymbol{A} \Longrightarrow \nonumber
\end{equation}
\begin{equation}
\boldsymbol{A} = [\boldsymbol{\mathcal{M}}(\mathcal{R})]^{-1}\boldsymbol{H}_{\text{interface}} \Longrightarrow \nonumber
\end{equation}
\begin{equation}
 \boldsymbol{{\cal H}}_{\text{outer}} =  \boldsymbol{\mathcal{M}}(\mathcal{R}_{\text{outer}})[\boldsymbol{\mathcal{M}}(\mathcal{R})]^{-1}\boldsymbol{H}_{\text{interface}}, \label{eq:matchingmachinery}  
\end{equation}
where 
\begin{equation}
\boldsymbol{M_1}(x) = \left( \begin{array}{ccccc} 1  & x^2 & x^4 & \log{x} & x^2 \log{x} \\ 0 & 2x & 4x^3 & 1/x & x + 2x \log{x} \\ 0 & 2 & 12 x^2 & -1/x^2 & 2\log{x} + 3 \\ 0 & 0 & 24x & 2/x^3 & 2/x \\ 0 & 0 & 24 & -6/x^4 & -2/x^2  \end{array} \right). 
\end{equation}

\subsection{Re-writing the Far-field Clamped Boundary Conditions}
Using (\ref{eq:matchingmachinery}), the boundary conditions at $\rho = \mathcal{R}_{outer}$ given in (\ref{eq:clampedbc}) can be written in the form
\begin{subequations}
\begin{eqnarray}
&&\frac{\mathcal{R}_{outer}^4}{64}\left( {\cal H}_4 + \frac{2 {\cal H}_3}{\mathcal{R}} - \frac{ {\cal H}_2}{\mathcal{R}^2} + \frac{{\cal H}_1}{\mathcal{R}^3} \right) \nonumber \\
&-&\frac{\mathcal{R}^2 \mathcal{R}^2_{\text{outer}}}{8}\log{\left( \frac{\mathcal{R}_{\text{outer}}}{\mathcal{R}} \right)}\left( {\cal H}_4 - \frac{3 {\cal H}_2}{\mathcal{R}^2} + \frac{3 {\cal H}_1}{\mathcal{R}^3} \right) \nonumber \\
&=& \mathcal{O}\left( {\cal H}_0\left(\frac{\mathcal{R}_{\text{outer}}}{\mathcal{R}}\right)^2 \right),
\end{eqnarray}
and
\begin{eqnarray}
&-&\frac{\mathcal{R}^2 \mathcal{R}^2_{\text{outer}}}{8}\log{\left( \frac{\mathcal{R}_{\text{outer}}}{\mathcal{R}} \right)}\left( {\cal H}_4 - \frac{3 {\cal H}_2}{\mathcal{R}^2} + \frac{3 {\cal H}_1}{\mathcal{R}^3} \right) \nonumber \\
&=& \mathcal{O}\left( {\cal H}_0\left(\frac{\mathcal{R}_{\text{outer}}}{\mathcal{R}}\right)^2 \right).
\end{eqnarray}
These two conditions rearrange to give
\begin{equation}
  {\cal H}_4 - \frac{3 {\cal H}_2}{\mathcal{R}^2} + \frac{3 {\cal H}_1}{\mathcal{R}^3} = \mathcal{O}\left( \frac{{\cal H}_0}{\mathcal{R}^4 \log{\left(\mathcal{R}_{\text{outer}}/\mathcal{R}\right) }} \right),  
\end{equation}
\begin{equation}
{\cal H}_4 + \frac{2 {\cal H}_3}{\mathcal{R}} - \frac{ {\cal H}_2}{\mathcal{R}^2} + \frac{{\cal H}_1}{\mathcal{R}^3} = \mathcal{O}\left( \frac{{\cal H}_0}{\mathcal{R}^2\mathcal{R}_{\text{outer}}^2} \right).    
\end{equation}
Moving back to tensorial notation, we see that to leading order in $\mathcal{R}_{\text{outer}}$ the far field boundary conditions can be rewritten as the zero pressure condition
\begin{equation}
 \nabla^4 {\cal H}(\mathcal{R},\tau) = 0 + \mathcal{O}\left( \frac{{\cal H}_0}{\mathcal{R}^2\mathcal{R}_{\text{outer}}^2} \right), \label{eq:clampedfarfieldbc1}   
\end{equation}
together with the force free condition
\begin{equation}
    \frac{\partial}{\partial \rho}\left( \nabla^2 {\cal H}(\mathcal{R},\tau)\right) = 0 + \mathcal{O}\left( \frac{{\cal H}_0}{\mathcal{R}^4 \log{\left(\mathcal{R}_{\text{outer}}/\mathcal{R}\right) }} \right). \label{eq:clampedfarfieldbc2}    
\end{equation}
\end{subequations}

\subsection{Re-writing the Far-field Free Beam Boundary Conditions}
In the same way, using (\ref{eq:matchingmachinery}), the boundary conditions at $\rho = \mathcal{R}_{outer}$ given in (\ref{eq:notclampedbc}) can be written in the form
\begin{subequations}
\begin{eqnarray}
&&\frac{3\mathcal{R}_{outer}^4}{16}\left( {\cal H}_4 + \frac{2 {\cal H}_3}{\mathcal{R}} - \frac{ {\cal H}_2}{\mathcal{R}^2} + \frac{{\cal H}_1}{\mathcal{R}^3} \right) \nonumber \\
&-&\frac{\mathcal{R}^2 \mathcal{R}^2_{\text{outer}}}{4}\log{\left( \frac{\mathcal{R}_{\text{outer}}}{\mathcal{R}} \right)}\left( {\cal H}_4 - \frac{3 {\cal H}_2}{\mathcal{R}^2} + \frac{3 {\cal H}_1}{\mathcal{R}^3} \right) \nonumber \\
&=& \mathcal{O}\left( {\cal H}_0\left(\frac{\mathcal{R}_{\text{outer}}}{\mathcal{R}}\right)^2 \right),
\end{eqnarray}
and
\begin{eqnarray}
&-& 4\mathcal{R}^2 \mathcal{R}^2_{\text{outer}}\log{\left( \frac{\mathcal{R}_{\text{outer}}}{\mathcal{R}} \right)}\left( {\cal H}_4 - \frac{3 {\cal H}_2}{\mathcal{R}^2} + \frac{3 {\cal H}_1}{\mathcal{R}^3} \right) \nonumber \\
&=& \mathcal{O}\left( {\cal H}_0\left(\frac{\mathcal{R}_{\text{outer}}}{\mathcal{R}}\right)^2 \right).
\end{eqnarray}
These two conditions rearrange to give
\begin{equation}
  {\cal H}_4 - \frac{3 {\cal H}_2}{\mathcal{R}^2} + \frac{3 {\cal H}_1}{\mathcal{R}^3} = \mathcal{O}\left( \frac{{\cal H}_0}{\mathcal{R}^4 \log{\left(\mathcal{R}_{\text{outer}}/\mathcal{R}\right) }} \right),  
\end{equation}
\begin{equation}
{\cal H}_4 + \frac{2 {\cal H}_3}{\mathcal{R}} - \frac{ {\cal H}_2}{\mathcal{R}^2} + \frac{{\cal H}_1}{\mathcal{R}^3} = \mathcal{O}\left( \frac{{\cal H}_0}{\mathcal{R}^2\mathcal{R}_{\text{outer}}^2} \right).    
\end{equation}
Moving back to tensorial notation, we see that to leading order in $\mathcal{R}_{\text{outer}}$ the far field boundary conditions can be rewritten as the zero pressure condition
\begin{equation}
 \nabla^4 {\cal H}(\mathcal{R},\tau) = 0 + \mathcal{O}\left( \frac{{\cal H}_0}{\mathcal{R}^2\mathcal{R}_{\text{outer}}^2} \right), \label{eq:notclampedfarfieldbc1}   
\end{equation}
together with the force free condition
\begin{equation}
    \frac{\partial}{\partial \rho}\left( \nabla^2 {\cal H}(\mathcal{R},\tau)\right) = 0 + \mathcal{O}\left( \frac{{\cal H}_0}{\mathcal{R}^4 \log{\left(\mathcal{R}_{\text{outer}}/\mathcal{R}\right) }} \right). \label{eq:notclampedfarfieldbc2}
    \end{equation}
\end{subequations}
Noting that (\ref{eq:clampedfarfieldbc1}), (\ref{eq:clampedfarfieldbc2}) and (\ref{eq:notclampedfarfieldbc1}), (\ref{eq:notclampedfarfieldbc2}) are identical, we see that both set of boundary conditions at $\rho = \mathcal{R}_{\text{outer}}$, when rewritten in terms of derivatives of ${\cal H}$ at $\rho = \mathcal{R}$, give at leading order in $\mathcal{R}_{\text{outer}}$ the same conditions for ${\cal H}$.

\section{Mode Zero Similarity Solution} \label{modezerosimilaritysolution}
To make further analytic progress, we look for a similarity solution in the mode zero case i.e.
\begin{equation}
    b_2 = 0 \Longrightarrow \frac{\partial}{\partial \rho}\left( \nabla^4 \mathcal{H} \right) = 0 \Longrightarrow \nonumber
\end{equation}    
\begin{equation}    
    \mathcal{H} = F\left(A - \left(\frac{\rho}{\mathcal{R}}\right)^2\right),
\end{equation}
where we have utilised the horizontal boundary conditions for $\mathcal{H}$, $A$ is a constant and $F = F(\tau)$ is independent of $\rho$. Evaluating (\ref{eq:nondimengovern7}) at $\rho = \mathcal{R}$ then gives
\begin{equation}
    \frac{\partial \mathcal{H}}{\partial \tau}(\mathcal{R},\tau) + \frac{1}{\mathcal{R}}\frac{\partial}{\partial \rho}\left( \rho \mathcal{H}v_s \right) \Big{|}_{\mathcal{R}} = \mathcal{H}(\mathcal{R},\tau) \Longrightarrow \nonumber
\end{equation}
\begin{equation}
    (A - 1)\left[ \frac{\partial F}{\partial \tau} + \frac{2 F}{\mathcal{R}}\frac{\partial \mathcal{R}}{\partial \tau} - F \right] = 0 \Longrightarrow F = \frac{F_0 e^{\tau}}{\mathcal{R}^2}.
\end{equation}
Applying the initial condition ${\cal H}_0(\rho = 0, \tau = 0) = 1$ and defining the incline ratio $m_0$, a measure of the initial flatness of the biofilm, as satisfying
\begin{equation}
m_0 = \frac{h(r = R(0),t = 0)}{h(r = 0,t = 0)} = {\cal H}(\rho = 1, \tau = 0),
\end{equation}
we find $F_0 = 1 - m_0$ and $A = 1/(1 - m_0)$ i.e.
\begin{equation}
    \mathcal{H} = \frac{e^{\tau}}{\mathcal{R}^2}\left( 1  - (1 - m_0)\left( \frac{\rho}{\mathcal{R}} \right)^2\right).
\end{equation}

Defining $f = f(\rho, \, \tau) = \mathcal{H} \varphi e^{\tau}$, the continuity equation (\ref{eq:nondimengovern7}) becomes
\begin{equation}
    \frac{\partial f}{\partial \tau} = - \left( \frac{1}{\mathcal{R}}\frac{\partial \mathcal{R}}{\partial \tau} \right)\frac{1}{\rho}\frac{\partial}{\partial \rho}\left( \rho^2 f \right).
\end{equation}
Looking for a similarity solution of the form $f = f_1(\mathcal{R})f_2(\eta)$ where $\eta = \rho / \mathcal{R}$, this simplifies to give
\begin{equation}
    \frac{\partial f_1}{\partial \mathcal{R}} = -\frac{2 f_1}{\mathcal{R}} \Longrightarrow f_1 = \frac{1}{\mathcal{R}^2} \Longrightarrow \varphi = \varphi_0(\eta),    
\end{equation}
Here $\varphi_0(\rho) = \varphi(\rho,\tau = 0)$ is set from the initial conditions ($\varphi_0$ must satisfy the properties $\varphi = \phi_{\infty}$ and $\partial \varphi / \partial \rho = 0$ at $\tau = 0$).

 Finally, we seek an analytical solution for $\boldsymbol{\xi^{*}} = (\xi^*, \, \zeta^{*})$ with minimal $z^{*}$ dependence. Setting $a_2 = 0$, (\ref{eq:xifunctionalform}) and (\ref{eq:zetafunctionalform}) simplify to become 
\begin{equation}
     \zeta^{*} = a_0 + z^{*} \left( 1 + \frac{A}{\mathcal{H}} \right), \quad \xi^{*} = b_0 + \frac{B z^{*}}{\mathcal{H}},
\end{equation} 
where $\{ a_0, \, b_0, \, A, \, B \}$ are all independent of $\tau$. (\ref{eq:stiffgoverning1}) is automatically satisfied. (\ref{eq:stiffgoverning2}) and (\ref{eq:stiffgoverning3}) reduce to 
\begin{equation}
    \frac{\partial}{\partial \rho}\left( \frac{\rho B}{\mathcal{H}} \right) = 0 \Longrightarrow \frac{\rho B(\rho)}{F(\tau)\left( A - \left(\rho / \mathcal{R} \right)^2 \right)} \Longrightarrow B = 0.
\end{equation}
\begin{equation}
    \frac{6 \rho}{\mathcal{R} \mathcal{H}} \frac{\partial \mathcal{R}}{\partial \tau} \left( -\frac{A}{\mathcal{H}}\left( 1 - \frac{\partial b_0}{\partial \rho} \right) \right) = 0 \Longrightarrow
\end{equation}    
\begin{equation}    
    \frac{\partial b_0}{\partial \rho} = 1 \Longrightarrow b_0 = \xi_0 + \rho, 
\end{equation}
where $\xi_0$ is a constant set from the initial conditions. Finally, we set for simplicity $A$ to a constant $\zeta_0$. Hence, the stress boundary condition (\ref{eq:pressurecondition}1) simplifies to become 
\begin{equation}
    \frac{\zeta_0}{\mathcal{H}} + \frac{\tilde{K}}{\rho}\frac{\partial}{\partial \rho}\left( B_0 \rho + \rho^2 \right) = C_0 \text{ at } \rho = \mathcal{R}.
\end{equation}
Applying the initial condition $\mathcal{R}(0) = 1$, we recover the cubic equation which describes the evolution of $\mathcal{R}$
\begin{equation}
    e^{-\tau} \mathcal{R}^3 + \mathcal{R} \left( \Xi - 1 \right) - \Xi = 0, \label{eq:Rgoverningequation1}
\end{equation}
Here, the non-dimensional evolution constant $\Xi$ is 
 \begin{equation}
     \Xi = \frac{\tilde{K}\xi_0 m_0}{\zeta_0},
 \end{equation}
and is thus is determined from the initial conditions as the product of the incline ratio and a ratio between horizontal 
and vertical stresses. 

Now, Cardano's formula for depressed cubic equations states that for the equation 
\begin{equation}
x^3 + p x + q = 0,
\end{equation}
where $p$ and $q$ are real, if $\Lambda(p,q) = 4 p^3 + 27 q^2 > 0$ then the equation has the single real root
\begin{align}
    x &= \left( -\frac{q}{2} + \sqrt{\frac{q^2}{4} + \frac{p^3}{27} } \right)^{1/3} \nonumber \\
    &+ \left( -\frac{q}{2} - \sqrt{\frac{q^2}{4} + \frac{p^3}{27} } \right)^{1/3}, 
\end{align}
with the other two roots being complex conjugates. If $\Lambda < 0$ there are three real roots but they can 
not be represented by an algebraic expression involving only real numbers. This was called by 
Cardano the \textit{casus irreducibilis} (Latin for `the irreducible case'). 

For (\ref{eq:Rgoverningequation1}), we have
\begin{subequations}
\begin{align}
p &= e^{\tau}(\Xi - 1), \\
q &= - e^{\tau}\Xi, \\
\Lambda &=  e^{2\tau} \left( 4 e^{\tau} \left( \Xi - 1 \right)^3 + 27 \Xi^2 \right).
\end{align}
\end{subequations}
Hence, $\Lambda < 0$ when $\Xi < 1$ and $\tau$ satisfies
\begin{equation}
    \tau > \tau_{\text{crit}} = \log{\left( \frac{27 \Xi^2}{4(1 - \Xi)^3} \right)}.
\end{equation}
Thus, for general $\Xi$ and $\tau$, (\ref{eq:Rgoverningequation1}) does not admit an analytical solution. Instead, 
this cubic equation is solved numerically using the MATLAB inbuilt function fzero \cite{Brent73}. Since cubic equations 
have up to three real roots, we select the correct root by locating the root that is closest to the value for 
$\mathcal{R}$ found at the previous time step, noting that by definition $\mathcal{R}(\tau = 0) = 1$.

\section{Newtonian Model}
Here, for comparison, we analyze the corresponding mathematical model in which, as in Seminara et. al. \cite{Seminara12}, 
the intrinsic elasticity of the biofilm extracellular matrix is neglected. In this case, a solution with power law growth 
tending to a maximum finite biofilm radius is not supported, demonstrating that matrix elasticity is essential 
to capture the behavior we have observed experimentally.

\subsection{Dimensionless shallow-layer scalings}
In the same way as for the poroelastic model, we scale radial and vertical lengths with the initial 
radius $R_0 = R(t = 0)$ and height $H_0 = h(r = 0, t = 0)$ of the biofilm, respectively, permeability with the 
characteristic permeability scale $\kappa_0$, pressure with the vertical confinement pressure scale and time with thath
for biofilm growth i.e. $\{ r, R \} \sim R_0$, $\{ z, h \} \sim H_0$, $\kappa \sim \kappa_0$, $p \sim P_0 = B H_0 / R_0^4$ 
and $t \sim 1/g$. Hence, we find $\{ u_f, u_s \} \sim g R_0$ and $\{ w_f, w_s \} \sim g H_0$. Similarly, we denote the dimensionless form of a function $f$ by $f^{*}$ and set for clarity
\begin{equation}
    \rho = r^{*} = \frac{r}{R(0)}, \ {\cal H} = h^{*} = \frac{h(r,t)}{h(0,0)}, \ \tau = \tau^{*} = g t, \nonumber 
\end{equation}    
\begin{equation}    
    {\mathcal{R}} = R^{*} = \frac{R(t)}{R(0)}, \ \mathcal{P} = p^{*} = \frac{p}{P_0}. \nonumber
\end{equation}
As above, we assume that the biomass volume fraction $\phi$ is independent of $z^{*}$,
\begin{equation}
    \frac{\partial \phi}{\partial z^{*}} = 0. \label{eq:newtonianzindependentvolumefraction}
\end{equation}
In nondimensional form, the governing equations for this system become
\begin{subequations}
\begin{equation}
    \frac{\partial \phi}{\partial \tau} + \frac{1}{\rho}\frac{\partial}{\partial \rho}(\rho \phi u^{*}_s) + \frac{\partial}{\partial z^{*}}(\phi w_s^{*}) = \phi, \label{eq:newtoniangoverning1}
\end{equation}
\begin{equation}
    -\frac{\partial \phi}{\partial \tau} + \frac{1}{\rho}\frac{\partial}{\partial \rho}(\rho(1-\phi)u^{*}_f) + \frac{\partial}{\partial z^{*}}((1-\phi)w^{*}_f) = -\phi, \label{eq:newtoniangoverning2}
\end{equation}
\begin{equation}
    u_s^{*} - u_f^{*} = \frac{\epsilon^2}{W_1}\frac{\kappa^*}{1 - \phi}\frac{\partial \mathcal{P}}{\partial \rho}, \label{eq:newtoniangoverning3}
\end{equation}
\begin{equation}
    w_s^{*} - w_f^{*} = \frac{1}{W_1}\frac{\kappa^{*}}{1-\phi}\frac{\partial \mathcal{P}}{\partial z^{*}}, \label{eq:newtoniangoverning4}
\end{equation}
\begin{eqnarray}
 \frac{\epsilon^2}{W_1} \frac{\partial \mathcal{P}}{\partial \rho} = &\mu& \Bigg{(} \frac{\partial^2 u_s^{*}}{\partial {z^{*}}^2} \nonumber \\
&+& \left( \frac{H_0}{R_0} \right)^2\frac{1}{\rho}\frac{\partial}{\partial \rho}\left( \rho \frac{\partial u_s^{*}}{\partial \rho} \right)\Bigg{)}, \label{eq:newtoniangoverning5}
\end{eqnarray}
\begin{eqnarray}
 \frac{1}{W_1}\frac{\partial \mathcal{P}}{\partial z^{*}} = &\mu& \Bigg{(}\frac{\partial^2 w_s^{*}}{\partial {z^{*}}^2} \nonumber \\
&+& \left( \frac{H_0}{R_0} \right)^2\frac{1}{\rho}\frac{\partial}{\partial \rho}\left( \rho \frac{\partial w_s^{*}}{\partial \rho} \right)\Bigg{)}, \label{eq:newtoniangoverning6}
\end{eqnarray}
\end{subequations}
where the non-dimensional constants $\{ \mu, \, W_1 \}$ satisfy
\begin{equation}
    \mu = \frac{\kappa_0}{H_0^2}\frac{\mu_s}{\mu_f}, \quad W_1 = \frac{\mu_f g H_0^2}{\kappa_0 P_0}. \label{eq:viscosityscalinggroup}
\end{equation}
Here $W_1$, defined in (\ref{eq:nondimensionalconstants}), is a dimensionless measure of the ability of flow to generate a vertical pressure gradient, $\mu$ is the non-dimensional biofilm viscosity scaling group and $\mu_s$ is the dimensional Newtonian viscosity of the biofilm solid phase. Utilising the typical experimental values for the scalings 
together with $\mu_s \sim 10^2 \, \si{\pascal\second }$ we see that $ \{ W_1, \, \epsilon^{-2}W_1 \} \ll 1$ while $\mu \approx 2.8 = \mathcal{O}(1)$.
\subsection{Vertical boundary conditions}
As before, imposing no-slip boundary conditions at both the lower and upper boundaries gives
\begin{subequations}
\begin{equation}
    w_f^{*} = w_s^{*} = u_s^{*} \text{ at } z^{*} = 0, \label{eq:newtonianbc1}
\end{equation}
\begin{equation}
    u_s^{*} = u_f^{*} = 0 \, , \, w_f^{*} = w_s^{*} = \frac{\partial {\cal H}}{\partial t} \text{ at } z^{*} = {\cal H}. \label{eq:newtonianbc2}
\end{equation}
Balancing normal stress at the biofilm sheet interface gives
\begin{equation}
    {\mathcal{P}}\Big{|}_{z^{*} = {\cal H}} = {\mathcal{P}}_0 + \nabla^4 {\cal H} + 2\mu W_1\frac{\partial w_s^{*}}{\partial z^{*}}\Big{|}_{z^{*} = {\cal H}} \label{eq:newtonianpressurecondition}
\end{equation}
\end{subequations}
where ${\mathcal{P}}_0$ is a constant reference pressure. Working at leading order in $W_1$, combining (\ref{eq:newtoniangoverning4}) and (\ref{eq:newtonianpressurecondition}) gives
\begin{subequations}
\begin{equation}
    \frac{\partial \mathcal{P}}{\partial z^{*}} = 0 + \mathcal{O}\left( W_1 \right) \Longrightarrow \mathcal{P} = {\mathcal{P}}_0 + \nabla^4 {\cal H} + \mathcal{O}\left( W_1 \right).
\end{equation}
Hence, applying (\ref{eq:newtoniangoverning3}) gives the differential equation for $\mathcal{H}$
\begin{equation}
    \frac{\partial}{\partial \rho}\left( \nabla^4 \mathcal{H} \right) = 0 + \mathcal{O}(\epsilon^{-2} W_1). \label{eq:newtonianheighcondition}
\end{equation}
Similarly, combining ((\ref{eq:newtoniangoverning4}) and (\ref{eq:newtoniangoverning6})) and ((\ref{eq:newtoniangoverning4}) and (\ref{eq:newtoniangoverning6})) yields respectively 
\begin{equation}
    u_s^{*} - u_f^{*} = \mu \frac{\kappa^{*}}{1 - \phi}\frac{\partial^2 u_s^{*}}{\partial {z^{*}}^2} + \mathcal{O}\left( \epsilon^2 \right), 
\end{equation}
\begin{equation}    
    w_s^{*} - w_f^{*} = \mu \frac{\kappa^{*}}{1 - \phi}\frac{\partial^2 w_s^{*}}{\partial {z^{*}}^2} + \mathcal{O}\left( \epsilon^2 \right).
\end{equation}
Finally, integrating (\ref{eq:newtoniangoverning5}) using the boundary conditions given in (\ref{eq:newtonianbc1}) and (\ref{eq:newtonianbc2}) gives
\begin{equation}
    u_s^{*} = -\frac{z^{*}({\cal H}-z^{*})}{2\mu}\left( \frac{\epsilon^2}{W_1} \frac{\partial {\mathcal{P}}}{\partial \rho}\right) + \mathcal{O}(\epsilon^2). \label{eq:newtonianus}
\end{equation}
\end{subequations}

\subsection{Vertically averaged governing equations}
We denote vertically averaged quantities by triangular brackets, namely for an arbitrary function $f$ we define 
$\langle f \rangle = {\cal H}^{-1}\int^{\cal H}_{0} f \, dz^{*}$, and for clarity set 
\begin{equation}
    \varphi = \langle \phi \rangle, \ k = \langle \kappa \rangle, \ v_s = \langle u_s \rangle = -\frac{\mathcal{H}^2}{12\mu}\left( \frac{\epsilon^2}{W_1}\frac{\partial \mathcal{P}}{\partial \rho} \right) + \mathcal{O}(\epsilon^2). \nonumber
\end{equation}
Integrating (\ref{eq:newtoniangoverning1}) in the $z^{*}$ direction yields
\begin{equation}
      \frac{\partial}{\partial \tau}(\varphi {\cal H}) + \frac{1}{\rho}\frac{\partial}{\partial \rho}\left(\rho \langle\phi u_s^{*} \rangle {\cal H} \right) = \varphi{\cal H}. \label{eq:newtonianfullgoverning1}  
\end{equation}
Similarly, integrating (\ref{eq:newtoniangoverning1})+(\ref{eq:newtoniangoverning2}) in the $z^{*}$ direction gives the continuity equation
\begin{eqnarray}
 \frac{\partial {\cal H}}{\partial \tau} &=& \frac{1}{\rho}\frac{\partial}{\partial \rho}\left( \rho{\cal H} \left( \frac{k \epsilon^2}{W_1} \frac{\partial \mathcal{P}}{\partial \rho} - v_s \right) \right) \nonumber \\
&=& \frac{1}{\rho}\frac{\partial}{\partial \rho}\left( \rho {\cal H} \left( k + \frac{{\cal H}^2}{12 \mu} \right) \left( \frac{\epsilon^2}{W_1}\frac{\partial}{\partial \rho}\left( \nabla^4 {\cal H} \right) \right) \right) \nonumber \\
&+& \mathcal{O}(\epsilon^2)
 \label{eq:newtonianfullgoverning2}
\end{eqnarray}

\subsection{Vertically averaged boundary conditions} \label{newtonianhorizontalbc}
As in the poroelastic model, we have the boundary conditions
\begin{subequations}
\begin{equation}
    \frac{\partial {\cal H}}{\partial \rho} = \frac{\partial {\mathcal P}}{\partial \rho} = 0 \mbox{ at } \rho = 0, \label{eq:newtonianfullbc1}
\end{equation}
\begin{equation}
    \nabla^4 \mathcal{H} = \mathcal{R}^3\frac{\partial^3 {\cal H}}{\partial \rho^3} + \mathcal{R}\frac{\partial^2 {\cal H}}{\partial \rho^2} - \frac{\partial {\cal H}}{\partial \rho} = 0 \mbox{ at } \rho = \mathcal{R}, \label{eq:newtonianfullbc2}
\end{equation}
\begin{equation}
    \varphi = \varphi_{\infty} \mbox{ at } \rho = \mathcal{R}, \label{eq:newtonianfullbc3}
\end{equation}    
\begin{equation}    
    \frac{\partial \varphi}{\partial \tau} = 0 \mbox{ at } \rho = \mathcal{R}. \label{eq:newtonianfullbc4}
\end{equation}
\end{subequations}
Similarly, polymer volume conservation yields the evolution condition 
\begin{equation}
    \frac{\partial \mathcal{R}}{\partial t} = \frac{\langle \phi \, u_s^{*} \rangle }{\phi_{\infty}} = v_s \mbox{ at } \rho = \mathcal{R}(t). \label{eq:newtonianfullbc5}
\end{equation}
As above, (\ref{eq:newtonianheighcondition}) together with the boundary conditions in (\ref{eq:newtonianfullbc2}) admits the similarity solution
\begin{equation}
    \mathcal{H} = F_0 + F_1 \rho^2,
\end{equation}
where $F_0 = F_0(\tau)$ and $F_1 = F_1(\tau)$ are independent of $\rho$. Integrating (\ref{eq:newtonianfullgoverning2}) with respect to $\rho$ then gives
\begin{equation}
    \frac{\rho^2}{2}\frac{\partial F_0}{\partial \tau} + \frac{\rho^4}{4}\frac{\partial F_1}{\partial \tau} = -\frac{12 \rho \mu}{\mathcal{H}}\left( k + \frac{\mathcal{H}^2}{12 \mu} \right) \Longrightarrow \nonumber
\end{equation}    
\begin{equation}    
    v_s = - \frac{\mathcal{H}}{\rho \left( 12 \mu k + \mathcal{H}^2 \right)}\left( \frac{\rho^2}{2}\frac{\partial F_0}{\partial \tau} + \frac{\rho^4}{4}\frac{\partial F_1}{\partial \tau} \right). \label{eq:newtonianreducedvs}
\end{equation}
Here, we have used (\ref{eq:newtonianfullbc1}) to set the integration constant to 0. In general, one can not make further analytic progress.
\subsection{Finite radius solution}
Experimentally, we see that the radius of the biofilm tends to a finite value i.e. the system supports a biofilm with constant radius $\mathcal{R} = \mathcal{R}_{\infty}$. In this case, (\ref{eq:newtonianfullbc5}) simplifies to give
\begin{equation}
    v_s\big{|}_{\mathcal{R}_{\infty}} = 0 \Longrightarrow \frac{\partial F_0}{\partial \tau} + \frac{\mathcal{R}_{\infty}^2}{2}\frac{\partial F_1}{\partial \tau} = 0 \Longrightarrow \nonumber
\end{equation}    
\begin{equation}    
    F_0 = C_1 - \frac{\mathcal{R}_{\infty}^2 F_1}{2},
\end{equation}
where $C_1$ is a constant. Since $\mathcal{H} > 0 \, \forall \rho \in [0, \mathcal{R}_{\infty}]$, evaluating $\mathcal{H}$ at $\rho = 0$ and $\rho = \mathcal{R}_{\infty}$ gives
\begin{equation}
    \mathcal{H} = C_1 + F_1 \left( \rho^2 - \frac{\mathcal{R}_{\infty}^2}{2} \right) \Longrightarrow \nonumber
\end{equation}    
\begin{equation}    
    \left\{ \mathcal{H}\big{|}_{0} = C_1 - \frac{F_1 \mathcal{R}_{\infty}^2}{2}, \, \mathcal{H}\big{|}_{\mathcal{R}_{\infty}} = C_1 + \frac{F_1 \mathcal{R}_{\infty}^2}{2} \right\},
\end{equation}
namely $C_1$ is positive with the lower bound $C_1 > \mathcal{R}_{\infty}^2|F_1|/2$. Similarly, differentiating (\ref{eq:newtonianreducedvs}) with respect to $\rho$ at $\rho = \mathcal{R}_{\infty}$ gives
\begin{eqnarray}
    \frac{\partial v_s}{\partial \rho}\Big{|}_{\mathcal{R}_{\infty}} &=& -\frac{\mathcal{H}}{12 \mu k + \mathcal{H}^2}\left( \frac{\partial F_0}{\partial \tau} + \rho^2 \frac{\partial F_1}{\partial \tau} \right) \nonumber \\
    &=& -\frac{\mathcal{H}\mathcal{R}_{\infty}^2}{2 \left( \mathcal{H}^2 + 12 \mu k \right)}\frac{\partial F_1}{\partial \tau}.
\end{eqnarray}
Hence, evaluating (\ref{eq:newtonianfullgoverning1}) at $\rho = \mathcal{R}$, utilising the boundary conditions given above gives
\begin{equation}
\frac{\partial \mathcal{H}}{\partial \tau}\Bigg{|}_{\mathcal{R}_{\infty}} + \left( \mathcal{H}\frac{\partial v_s}{\partial \rho} \right) \Bigg{|}_{\mathcal{R}_{\infty}} = \mathcal{H}\Big{|}_{\mathcal{R}_{\infty}} \Longrightarrow \nonumber
\end{equation}
\begin{equation}
C_2 \frac{\partial F_1}{\partial \tau} = (F_1 + C_3)\left(C_2 + (F_1 + C_3)^2 \right) \Longrightarrow \nonumber
\end{equation}
\begin{eqnarray}
    \tau &=& \int \frac{C_2}{(F_1 + C_3)\left(C_2 + (F_1 + C_3)^2 \right)} \, dF_1 \nonumber \\
    &=& \int \frac{1}{F_1 + C_3} - \frac{(F_1 + C_3)}{C_2 + (F_1 + C_3)^2} \, dF_1 \nonumber \\
    &=& \ln{(F_1 + C_3)} - \frac{1}{2}\ln{\left( C_2 + (F_1 + C_3)^2 \right)} - \frac{1}{2}\ln{(\tilde{C})} \nonumber \\
    &=& \frac{1}{2}\ln{\left( \frac{(F_1 + C_3)^2}{\tilde{C} \left(C_2 + (F_1 + C_3)^2\right)} \right)} \Longrightarrow \nonumber 
\end{eqnarray}
\begin{equation}    
    (F_1 + C_3)^2 = \frac{\tilde{C}C_2 e^{2\tau}}{1 - \tilde{C}e^{2\tau}}, \label{eq:newtonianfailedsolution}
\end{equation}
where $\tilde{C}$ is a constant of integration and the positive constants $C_2$ and $C_3$ satisfy
\begin{equation}
    C_2 = \frac{48 \mu}{\mathcal{R}_{\infty}^4}k(\varphi_{\infty}), \quad C_3 = \frac{2 C_1}{\mathcal{R}_{\infty}^2}.
\end{equation}
Since the right hand side of (\ref{eq:newtonianfailedsolution}) is non-negative for all $\tau$, $\tilde{C} = 0$ and thus $F_1 = - C_3$. However, this then gives $\mathcal{H} = 0$ at $\rho = \mathcal{R}_{\infty}$ which is a contradiction. Hence, the Newtonian model does not support a constant radius solution and thus does not agree with experiments.


\begin{thebibliography}{0}%
\makeatletter
\providecommand \@ifxundefined [1]{%
 \@ifx{#1\undefined}
}%
\providecommand \@ifnum [1]{%
 \ifnum #1\expandafter \@firstoftwo
 \else \expandafter \@secondoftwo
 \fi
}%
\providecommand \@ifx [1]{%
 \ifx #1\expandafter \@firstoftwo
 \else \expandafter \@secondoftwo
 \fi
}%
\providecommand \natexlab [1]{#1}%
\providecommand \enquote  [1]{``#1''}%
\providecommand \bibnamefont  [1]{#1}%
\providecommand \bibfnamefont [1]{#1}%
\providecommand \citenamefont [1]{#1}%
\providecommand \href@noop [0]{\@secondoftwo}%
\providecommand \href [0]{\begingroup \@sanitize@url \@href}%
\providecommand \@href[1]{\@@startlink{#1}\@@href}%
\providecommand \@@href[1]{\endgroup#1\@@endlink}%
\providecommand \@sanitize@url [0]{\catcode `\\12\catcode `\$12\catcode
  `\&12\catcode `\#12\catcode `\^12\catcode `\_12\catcode `\%12\relax}%
\providecommand \@@startlink[1]{}%
\providecommand \@@endlink[0]{}%
\providecommand \url  [0]{\begingroup\@sanitize@url \@url }%
\providecommand \@url [1]{\endgroup\@href {#1}{\urlprefix }}%
\providecommand \urlprefix  [0]{URL }%
\providecommand \Eprint [0]{\href }%
\providecommand \doibase [0]{http://dx.doi.org/}%
\providecommand \selectlanguage [0]{\@gobble}%
\providecommand \bibinfo  [0]{\@secondoftwo}%
\providecommand \bibfield  [0]{\@secondoftwo}%
\providecommand \translation [1]{[#1]}%
\providecommand \BibitemOpen [0]{}%
\providecommand \bibitemStop [0]{}%
\providecommand \bibitemNoStop [0]{.\EOS\space}%
\providecommand \EOS [0]{\spacefactor3000\relax}%
\providecommand \BibitemShut  [1]{\csname bibitem#1\endcsname}%
\let\auto@bib@innerbib\@empty
\end{thebibliography}%


\begin{thebibliography}{99}

\bibitem{Bjarnsholt11} T. Bjarnsholt, Introduction to biofilms, in {\it Biofilm Infections}, 
T. Bjarnsholt, P.{\O}. Jensen, C. Moser and N. H{\o}iby, eds. (Springer, New York, 2011), pp. 1-9.

\bibitem{Hooke1665} R. Hooke,
Micrographia, or, Some physiological descriptions of minute bodies made by magnifying glasses :with observations and inquiries thereupon, \href{https://doi.org/10.5962/bhl.title.904}{London : Printed by J. Martyn and J. Allestry, printers to the Royal Society (1665)} 

\bibitem{Saraf84} P. G.Saraf, A. T. K. Cockett, Marcello Malpighi — A Tribute, \href{https://doi.org/10.1016/0090-4295(84)90087-6}{Urology {\bf 23}, 619-623 (1984)}

\bibitem{Wimpenny2000} J. Wimpenny, W. Manz, U. Szewzyk, Heterogeneity in biofilms, \href{https://doi.org/10.1111/j.1574-6976.2000.tb00565.x}{FEMS Microbiol. Rev. {\bf 24}, 661 (2000)}. 

\bibitem{vanL} A. van Leeuwenhoek, 
An abstract of a letter from Mr. Anthony Leewenhoeck at Delft, dated Sep. 17. 1683. Containing some microscopical observations, 
about animals in the scurf of the teeth, the substance call'd worms in the nose, the cuticula consisting of scales,
\href{https://doi.org/10.1098/rstl.1684.0030}{Phil. Trans. R. Soc. {\bf 14}, 568 (1684)}.

\bibitem{Costerton1994} J. W. Costerton, Z. Lewandowski, D. Debeer, D. Caldwell, D. Korber, G. James, Biofilms, the Customized Microniche, \href{https://doi.org/10.1128/jb.176.8.2137-2142.1994}{J Bacteriol. {\bf 176}, 2137 (1994)}.

\bibitem{OToole2000} G. O’Toole, H. B. Kaplan, R. Kolter, Biofilm formation as microbial development, \href{https://doi.org/10.1146/annurev.micro.54.1.49}{Annu. Rev. Microbiol. {\bf 54}, 49-79 (2000)} 

\bibitem{Lopez10} D. L\'opez, H. Vlamakis, and R. Kolter, Biofilms,
\href{https://doi.org/10.1101/cshperspect.a000398}{Cold Spring Harb. Perspect. Biol. {\bf 2}, a000398 (2010)}.

\bibitem{Seminara12} A. Seminara, T.E.  Angelini. J.N. Wilking, H. Vlamakis, S. Ebrahim, R. Kolter, D.A. Weitz,
and M. P. Brenner, Osmotic spreading of {\it Bacillus subtilis} biofilms driven by an extracellular matrix,
\href{https://doi.org/10.1073/pnas.1109261108}{Proc. Natl. Acad. Sci. USA {\bf 109}, 1116 (2012)}.

\bibitem{Tam19} A. Tam, E.F. Green, S. Balasuriya, E.L. Tek, J.M. Gardner, J.F. Sundstrom, V. Jiranek, and B.J. Binder,
A thin-film extensional flow model for biofilm expansion by sliding motility,
\href{https://doi.org/10.1098/rspa.2019.0175}{Proc. R. Soc. A {\bf 475}, 20190175 (2019)}.

\bibitem{Srinivasan19} S. Srinivasan, C.N. Kaplan, and L. Mahadevan,
A multiphase theory for spreading microbial swarms and films,
\href{https://doi.org/10.7554/eLife.42697}{eLife {\bf 8}, e42697 (2019)}.

\bibitem{Martinez-Corral19} R. Martinez-Corral, J. Liu, A. Prindle, G. M. S\"uel, and J. Garcia-Ojalvo,
Metabolic basis of brain-like electrical signalling in bacterial communities,
\href{https://doi.org/10.1098/rstb.2018.0382}{Phil. Trans. R. Soc. B {\bf 374}, 20180382 (2019)}.

\bibitem{Kempf19} F. Kempf, R. Mueller, E. Frey, and J. M. Yeomans, Active matter invasion,
\href{https://doi.org/10.1039/C9SM01210A}{Soft Matter {\bf 15}, 7538 (2019)}.

\bibitem{Conrad18} J. C. Conrad and R. Poling-Skutvik, Confined Flow: Consequences and implications for bacteria and biofilms,
\href{https://doi.org/10.1146/annurev-chembioeng-060817-084006}{Annu. Rev. Chem. Biomol. Eng. {\bf 9}, 175 (2018)}.

\bibitem{Khatoon18} Z. Khatoon, C.D. McTiernan, E.J. Suuronen, T.-F. Mah, and E. I. Alarcon,
Bacterial biofilms formation on implantable devices and approaches to its treatment and prevention,
\href{https://doi.org/10.1016/j.heliyon.2018.e01067}{Heliyon {\bf 4}, 1 (2018)}.

\bibitem{Oppenheimer13} Y. Oppenheimer-Shaanan, N. Steinberg, and I. Kolodkin-Gal,
Small molecules are natural triggers for the disassembly of biofilms,
\href{https://doi.org/10.1016/j.tim.2013.08.005}{Trends Microbiol. {\bf 21}, 594 (2013)}.

\bibitem{Ciofu17} O. Ciofu, E. Rojo-Molinero, M.D. Maci\`a, and A. Oliver,
Antibiotic treatment of biofilm infections,
\href{https://doi.org/10.1111/apm.12673}{APMIS {\bf 125}, 304 (2017)}.

\bibitem{Stewart02} P. S. Stewart, Mechanisms of antibiotic resistance in bacterial biofilms,
\href{https://doi.org/10.1078/1438-4221-00196}{Int. J. Med. Microbiol. {\bf 292}, 107 (2002)}.

\bibitem{Arciola18} C. R. Arciola, D. Campoccia, and L. Montanaro, 
Implant infections: adhesion, biofilm formation and immune evasion,
\href{https://doi.org/10.1038/s41579-018-0019-y}{Nat. Rev. Microbiol. {\bf 16}, 397 (2018)}.

\bibitem{Nowakowska14} J. Nowakowska, R. Landmann, and N. Khanna, 
Foreign body infection models to study host-pathogen response and antimicrobial tolerance of bacterial biofilms,
\href{https://doi.org/10.3390/antibiotics3030378}{Antibiotics {\bf 3}, 378 (2014)}.

\bibitem{Peppin05} S.S.L. Peppin, J.A.W. Elliott, and M. G. Worster,
Pressure and relative motion in colloidal suspensions,
\href{https://doi.org/10.1063/1.1915027}{Phys. Fluids {\bf 17}, 053301 (2005)}.

\bibitem{Wang01} H.F. Wang, {\it Theory of Linear Poroelasticity with Applications to Geomechanics and Hydrogeology}
(Princeton University Press, Princeton, 2001).

\bibitem{Hewitt15} D.R. Hewitt, J.A. Neufeld, and N.J. Balmforth, Shallow, gravity-driven flow in a poro-elastic layer,
\href{https://doi.org/10.1017/jfm.2015.361}{J. Fluid Mech. {\bf 778}, 335 (2015)}.

\bibitem{Gibson70} R.E. Gibson, R.L. Schiffman, and S.L. Pu, 
Plane strain and axially symmetric consolidation of a clay layer on a smooth impervious base,
\href{https://doi.org/10.1093/qjmam/23.4.505}{Q. J. Mech. Appl. Maths {\bf 23}, 505 (1970)}.

\bibitem{Barry97} S.I. Barry, G.N. Mercer, and C. Zoppou,
Deformation and fluid flow due to a source in a poro-elastitc layer,
\href{https://doi.org/10.1016/S0307-904X(97)00097-8}{Appl. Math. Model. {\bf 21}, 681 (1997)}.

\bibitem{Charras09} G.T. Charras, T.J. Mitchison, and L. Mahadevan, Animal cell hydraulics,
\href{https://doi.org/10.1242/jcs.049262}{J. Cell Sci. {\bf 122}, 3233 (2009)}.

\bibitem{Flory53} P.J. Flory, {\it Principles of Polymer Chemistry} (Cornell University Press, Ithaca, 1953).

\bibitem{Winstanley11} H.F. Winstanley, M. Chapwanya, M.J. McGuinness, and A.C. Fowler,
A polymer-solvent model of biofilm growth,
\href{https://doi.org/10.1098/rspa.2010.0327}{Proc. R. Soc. A {\bf 467}, 1449 (2011)}.

\bibitem{pi_caveat} Note that due to the dead volume occupied by the bacteria cells, the extracellular matrix occupies a volume $\phi_m$ within a total volume of $1 - \phi$ and thus we expand in terms of $\phi_m / 1-\phi$ rather than $\phi_m$.

\bibitem{suppmat} See Supplemental Material at http://link.aps.org/supplemental/10.1103/XXX for further
theoretical details and experimental methods, which includes Refs. 31-36.

\bibitem{Liu15} J. Liu, A. Prindle, J. Humphries, M. Gabalda-Sagarra, D.D. Lee, S. Ly, J. Garcia-Ojalvo, and G. M. S\"uel,
Metabolic co-dependence gives rise to collective oscillations within biofilms,
\href{https://doi.org/10.1038/nature14660}{Nature {\bf 523}, 550 (2015)}.

\bibitem{Humphries17} J. Humphries, L. Xiong, J. Liu, A. Prindle, F. Yuan, H.A. Arjes, L. Tsimring, and G.M. S\"uel,
Species-independent attraction to biofilms through electrical signaling,
\href{https://doi.org/10.1016/j.cell.2016.12.014}{Cell {\bf 168}, 200 (2017)}.

\bibitem{Schindelin12} J. Schindelin, I. Arganda-Carreras, E. Frise, V. Kaynig, M. Longair, T. Pietzsch, 
S. Preibisch, C. Rueden, S. Saalfeld, B. Schmid, J.-Y. Tinevez, D.J. White, V. Hartenstein, K. Eliceiri, P. Tomancak, 
and A. Cardona, Fiji: an open-source platform for biological-image analysis,
\href{https://doi.org/10.1038/nmeth.2019}{Nature Methods {\bf 9}, 676 (2012)}.

\bibitem{Brent73} R. P. Brent, { \it Algorithms for Minimization Without Derivatives}
(Prentice-Hall Inc. 1973).

\bibitem{Picioreanu18} C. Picioreanu, F. Blauert, H. Horn, and M. Wagner,
Determination of mechanical properties of biofilms by modelling the deformation using optical coherence tomography,
\href{https://doi.org/10.1016/j.watres.2018.08.070}{Water Res. {\bf 145}, 588-598 (2018)}.

\bibitem{Ismail09} A.E. Ismail, G.S. Grest, D.R. Heine, and M. J. Stevens,
Interfacial Structure and Dynamics of Siloxane Systems: PDMS-Vapor and PDMS-Water,
\href{https://doi.org/10.1021/ma802805y}{Macromolecules {\bf 42}, 3186-3194 (2009)}.

\end{thebibliography}

\begin{thebibliography}{99}

\bibitem{Seminara12} A. Seminara, T.E.  Angelini. J.N. Wilking, H. Vlamakis, S. Ebrahim, R. Kolter, D.A. Weitz,
and M. P. Brenner, Osmotic spreading of {\it Bacillus subtilis} biofilms driven by an extracellular matrix,
\href{https://doi.org/10.1073/pnas.1109261108}{Proc. Natl. Acad. Sci. USA {\bf 109}, 1116 (2012)}.

\bibitem{Martinez-Corral19} R. Martinez-Corral, J. Liu, A. Prindle, G. M. S\"uel, and J. Garcia-Ojalvo,
Metabolic basis of brain-like electrical signalling in bacterial communities,
\href{https://doi.org/10.1098/rstb.2018.0382}{Phil. Trans. R. Soc. B {\bf 374}, 20180382 (2019)}.

\bibitem{Liu15} J. Liu, A. Prindle, J. Humphries, M. Gabalda-Sagarra, D.D. Lee, S. Ly, J. Garcia-Ojalvo, and G. M. S\"uel,
Metabolic co-dependence gives rise to collective oscillations within biofilms,
\href{https://doi.org/10.1038/nature14660}{Nature {\bf 523}, 550 (2015)}.

\bibitem{Humphries17} J. Humphries, L. Xiong, J. Liu, A. Prindle, F. Yuan, H.A. Arjes, L. Tsimring, and G.M. S\"uel,
Species-independent attraction to biofilms through electrical signaling,
\href{https://doi.org/10.1016/j.cell.2016.12.014}{Cell {\bf 168}, 200 (2017)}.

\bibitem{Schindelin12} J. Schindelin, I. Arganda-Carreras, E. Frise, V. Kaynig, M. Longair, T. Pietzsch, 
S. Preibisch, C. Rueden, S. Saalfeld, B. Schmid, J.-Y. Tinevez, D.J. White, V. Hartenstein, K. Eliceiri, P. Tomancak, 
and A. Cardona, Fiji: an open-source platform for biological-image analysis,
\href{https://doi.org/10.1038/nmeth.2019}{Nature Methods {\bf 9}, 676 (2012)}.

\bibitem{Brent73} R. P. Brent, { \it Algorithms for Minimization Without Derivatives}
(Prentice-Hall Inc. 1973).

\bibitem{Picioreanu18} C. Picioreanu, F. Blauert, H. Horn, and M. Wagner,
Determination of mechanical properties of biofilms by modelling the deformation using optical coherence tomography,
\href{https://doi.org/10.1016/j.watres.2018.08.070}{Water Res. {\bf 145}, 588-598 (2018)}.

\bibitem{Ismail09} A.E. Ismail, G.S. Grest, D.R. Heine, and M. J. Stevens,
Interfacial Structure and Dynamics of Siloxane Systems: PDMS-Vapor and PDMS-Water,
\href{https://doi.org/10.1021/ma802805y}{Macromolecules {\bf 42}, 3186-3194 (2009)}.


\end{thebibliography}
\end{document}